\documentclass[sn-basic,Numbered,pdflatex]{sn-jnl}

\usepackage{xurl}
\usepackage{hyperref}
\usepackage{enumerate}
\usepackage[group-minimum-digits=4,group-separator={,}]{siunitx}
\usepackage{graphicx}
\graphicspath{{figs/used/}}
\usepackage{amsmath}
\usepackage{amssymb}
\usepackage{physics}
\usepackage[capitalise,nameinlink]{cleveref}
\usepackage[labelfont=bf]{caption}

\begin{document}

\title{
  When Dialects Collide: How Socioeconomic Mixing Affects Language Use
}

\author*[1,2]{\fnm{Thomas} \sur{Louf}}\email{tlouf@fbk.eu}

\author[1]{\fnm{Jos\'e J.} \sur{Ramasco}}

\author[1]{\fnm{David} \sur{S\'{a}nchez}}

\author*[3,4]{\fnm{M\'arton} \sur{Karsai}}\email{mkarsai@renyi.hu}

\affil[1]{\orgname{Institute for Cross-Disciplinary Physics and Complex Systems IFISC (UIB-CSIC)}, \orgaddress{\city{Palma de Mallorca}, \country{Spain}}}
\affil[2]{\orgname{Fondazione Bruno Kessler}, \orgaddress{\city{Povo (TN)}, \country{Italy}}}
\affil[3]{\orgname{Department of Network and Data Science, Central European University}, \orgaddress{\city{Vienna}, \country{Austria}}}
\affil[4]{\orgname{National Laboratory for Health Security, HUN-REN Alfr\'ed R\'enyi Institute of Mathematics}, \orgaddress{\city{Budapest}, \country{Hungary}}}

\abstract{The socioeconomic background of people and how they use standard forms of language are not independent, as demonstrated in various sociolinguistic studies.
However, the extent to which these correlations may be influenced by the mixing of people from different socioeconomic classes remains relatively unexplored from a quantitative perspective.
In this work we leverage geotagged tweets and transferable computational methods to map deviations from standard English across eight UK metropolitan areas.
We combine these data with high-resolution income maps to assign a proxy socioeconomic indicator to home-located users. Strikingly, we find a consistent pattern suggesting that the more different socioeconomic classes mix, the less interdependent the frequency of their departures from standard grammar and their income become. Further, we propose an agent-based model of linguistic variety adoption that sheds light on the mechanisms that produce the observations seen in the data.
}


\maketitle

\section{Introduction}
As a primary means of communication, language shapes social interactions, and in return it is shaped by society and its many aspects.
Due to this feedback, one may naturally expect to see social differences between individuals reflected in the language they use.
Culture, politics, education, or the economy: all of these spheres of society can be decisive in shaping language and its variety.
In particular, it was early shown in sociolinguistics that the socioeconomic status (SES) is an important determinant of linguistic variation \cite{LabovSocialStratification1966}.
For instance, higher-status groups systematically favour standard forms, particularly in formal contexts, while lower-status speakers exhibit greater use of non-standard or vernacular variants.
The standard variety of language is codified in dictionaries and grammar books \cite{TrudgillIntroducingLanguage1992} and is most often put forward by a society's major institutions, following a given language ideology \cite{TrudgillStandardEnglish1999}.
In fact, standard language proficiency is a form of linguistic capital that provides social and economic advantages, reinforcing existing power structures \cite{BourdieuLanguageSymbolic2009}.
Empirical research across various linguistic communities confirms this relation between the standard variety and SES \cite{TrudgillSocialDifferentiation1974,EckertLinguisticVariation2000}.
However, recent studies challenge simplistic class-based models by highlighting the covert prestige of non-standard varieties \cite{TrudgillSexCovert1972} and the impact of globalization on linguistic inequality \cite{BlommaertSociolinguisticsGlobalization2010}.
These findings suggest that while SES remains a crucial factor in language variation, its effects are mediated by broader sociocultural dynamics.
In this work, we show that the interaction strength between socioeconomic classes is crucial to understand the patterns of standard language use observed in cities.

The PISA reports of the OECD~\cite{OECDWhereAll2019} consistently show in a quantified manner that students with lower socioeconomic background tend to have a lower reading proficiency.
While these reports confirm there is an issue to tackle, they are not extensive enough.
They do not test language production, and not the whole population but only a sample of students of a specific age.
Alternative empirical works are thus needed to understand better the phenomenon.
Data from social media have repeatedly proven useful to link SES and different social behaviours~\cite{GaoComputationalSocioeconomics2019}.
In particular, Twitter data were used in the past to study linguistic features and their variation, whether lexical~\cite{EisensteinDiffusionLexical2014,GoncalvesCrowdsourcingDialect2014,BokanyiRaceReligion2016,DonosoDialectometricAnalysis2017,GrieveMappingLexical2019,AlshaabiStorywranglerMassive2021,LoufAmericanCultural2023}, semantic~\cite{ScepanovicSemanticHomophily2017,MartincLeveragingContextual2020}, or in spelling~\cite{EisensteinSystematicPatterning2015,JorgensenChallengesStudying2015,GoncalvesMappingAmericanization2018}.
These dimensions of variation can even be mapped geographically, leveraging the geotags that users may attach to their tweets.
Therefore, geotagged Twitter corpora enable researchers to relate linguistic variation to a given social variable characterising the populations of the areas under study.
Tweets, as instances of written language production, may not be a priori seen as the most adapted data to study dialectal variation, often more strongly marked in casual oral language. % TODO: citation?
But in fact, previous research has found a high prevalence of conversational language on Twitter \cite{JavaWhyWe2007,GouwsContextualBearing2011,BrodyCooooooooooooooollllllllllllllUsing2011, EisensteinWhatBad2013}.
Still, this criticism might be valid when it comes to drawing absolute conclusions about the dialect used by single individuals from their tweets.
Also, the demographic biases that this source presents~\cite{MisloveUnderstandingDemographics2011} warrant caution, especially as they are stronger for users geotagging their tweets~\cite{PavalanathanConfoundsConsequences2015}.
This sample population tends to be younger than the general population, for instance.
Since age can also be strongly correlated with SES, the language observed through this lens could therefore be that of a relatively low social class.
But these issues are much less relevant when making comparisons between areas based on the tweets sent out by a statistically relevant number of residents, which is what we do in this work.
Indeed, we do not make the assumption that the language of the users included in our dataset is absolutely representative of the ``usual'' language of all actual residents of these areas, but rather that if there is geographical variation in the offline world, then it should be reflected on Twitter.
Furthermore, since all the conclusions presented in this work are based on analyses carried out at the level of single metropolitan areas, any wider geographical bias, such as the fact that Twitter users tend to be more urban, does not apply.
Moreover, one should not dismiss the two main advantages this data source offers.
First, as opposed to more traditional sociolinguistic methods, such as conducting interviews, social media corpora greatly alleviate the observer's paradox \cite{LabovPrinciplesLinguistic1972}, that is, the fact that sociolinguistic studies must ensure that people should not feel observed in the course of their language production.
Second is the unprecedented large amount of linguistic data available in a Twitter corpus, as it enables the study of thousands, if not hundreds of thousands or millions of individuals.
That is why, all in all, this data source has a great potential to explore the complex interplay between language variation and SES.
A first step was taken in \cite{AbitbolSocioeconomicDependencies2018}, where the authors analyse a few markers of non-standard language in France and their socioeconomic dependencies.
Here, we perform a two-fold approach, both empirical and theoretical, to investigate how Twitter users in eight metropolitan areas of the UK abide by the rules of the standard variety of English.

\section{Data preprocessing}

We leverage a database containing slightly more than $550$ million geotagged tweets posted from Great Britain by around $4.5$ million users between 2015 and 2021.
These tweets were collected through the filtered stream endpoint of the Twitter application programming interface.
As we are interested in natural language production, we start our analysis by filtering out users whose behaviour resembles that of a bot.
We first eliminate those tweeting at an inhuman rate, set at an average of ten tweets per hour over their whole tweeting period.
Then, we retain only those who tweeted either from a Twitter official app, Instagram, Foursquare or Tweetbot --- a popular third-party app.
These were selected because they are significantly popular among real users.
Also, consecutive geolocations implying speeds higher than a plane's (\SI{1000}{\kilo \meter \per \hour}) are detected to discard users.
Finally, in order to only keep residents of England and Wales, we impose that users must have tweeted from there in at least three consecutive months.
The values for the criteria given above were set as they were deemed reasonable and allowed us to safely discard problematic users.
To then assign an SES indicator to each user, we determine their area of residence from the geotags attached to their tweets.
Our unit areas for the study are the \SI{7201}{} middle layer super output areas (MSOAs), which are areas created by the Office for National Statistics of the UK for the output of the census estimates (Scotland and Northern Ireland are excluded from this census).
In the following, we will refer to them as cells to avoid using this acronym.
Each cell hosts at least \SI{5000}{} and at most \SI{15000}{} inhabitants, with a typical population of \SI{8000}{}.
Crucially, the average annual net income in these cells can be obtained from the census.
Sources for all census data at the MSOA level are provided in the Availability of data and materials section.
Using these data, we are thus able to couple Twitter users to a proxy indicator of their SES, by attributing them a cell of residence.
This indicator, that is income, obviously does not give a complete view of SES.
It can be complemented with a number of other dimensions, such as professional occupation, accumulated wealth, education or cultural capital.
But while it is limiting in this sense, it remains a simple, meaningful way to operationalize SES for the scope of our analysis.
An important caveat in the analysis is that tweets' geotags can be given at different geographical levels, though.
Some tweets include coordinates, which were more abundant prior to 2015, when the network changed their geolocation policy.
From the end of 2015 on, it has become more common to have geotags at the level of places.
Places are geographical areas that may range from full countries, to regions or provinces, cities, neighbourhoods or even points of interest.
Since the sizes of these places span orders of magnitude, some may intersect many cells.
There can therefore be so much uncertainty in the actual geographic origin of the tweet that it is preferable to discard it.
Our criterion here is that when the top four cells which contain most of the place's area do not contain more than \SI{90}{\percent} of its total area when put together, then the place, and all the tweets assigned to it, are discarded.
Now, to explain the heuristics we defined for residence attribution, let us first formalise some notation.
The tweets of a user $u$ attributed to a cell $c$ are the sum of those with coordinates and those with a place intersecting $c$.
If the tweets' place intersects several cells, we apply the following criterion to calculate the contribution to $c$:
\begin{equation}
r_{c} = \sum_p n_{ p} \frac{A_{p \cap c}}{A_p},
\end{equation}
where $n_p$ are the tweets in place $p$, $A_p$ is the area of $p$ and $A_{p \cap c}$ is the area of the intersection between $p$ and $c$.
The cell of residence of each user is thus the one from which they tweet more often between 6pm and 8am.
Besides, we also impose that the user must have tweeted at least three times and at least \SI{10}{\percent} of the time from that cell, and that at night the majority of their tweets were from there.
This heuristic is based on the results from \cite{LinInferringHome2018}, where they found that the majority of tweets sent out within this time range were actually sent from the home location of users.
It is also commonly used in mobility studies concerned with commuting patterns \cite{LouailMobilePhone2014}, which also validated the approach \cite{PhithakkitnukoonSocioGeographyHuman2012}.
All users for whom a cell of residence could not be attributed are subsequently discarded from the analysis.
We show in Fig.~S1 a population density map of Twitter users we obtained through this pipeline for the whole country and the analysed metropolitan areas. %\cref{SIfig:EW_twitter_pop}

The second ingredient we need for our analysis is a measurement of people's propensity to deviate from the features that define standard British English.
We quantify the usage of non-standard features using LanguageTool, an open-source grammar, style, and spell checker.
LanguageTool supports multiple national standards (e.g., British vs American English).
Further variation is not explicitly included.
Naturally, we select British English in this analysis.
There are several advantages of using such a tool over a pre-defined set of rules, as in~\cite{AbitbolSocioeconomicDependencies2018}.
First, this tool covers a very wide spectrum of potential non-standard features: LanguageTool has more than \SI{5500}{} rules defined for the English language.
These rules are split into 11 categories which include grammar, confused words or typos, for instance.
In this work we focus on the grammatical category, since standard English is predominantly distinguished from other varieties by its syntactic forms \cite{TrudgillStandardEnglish1999}.
In fact, LanguageTool can flag variation from the selected standard as incorrect (e.g., orthographic variants).
Specifically targetting the grammar category therefore also allows us to exclude performance errors such as typos (see Table S1).
However, as a standard-based tool, it can still potentially misclassify dialectal norms in specific cases, even when we focus on this category.
LanguageTool is nonetheless generally efficient for the purposes of our work.
Another important advantage of the tool is that it is implemented in $15$ languages.
Our study could thus quite easily be replicated in other countries.
After running the tool on our corpus, we find the grammatical category to be among the most common in our dataset (see Table~S1 of the Supplementary Material). %  \cref{SItab:cats_global_stats}
Within that category, we provide the ten most common departures from the standard variety found in Twitter in Table~S2. %\cref{SItab:top_grammar_mistakes}.
In the following, for the sake of brevity, we will use the phrase ``non-standard features'' to refer to non-standard grammar features, which we observe in practice using the LanguageTool detection.
Since we are interested in natural language production, before applying the tool we need to clean text from URLs, mentions of other users and hashtags.
Those are the URLs, mentions of other users and hashtags.
It is not completely obvious that the latter should be discarded, though.
Hashtags are used on Twitter to aggregate tweets by topics.
It is an important feature of the website, whose aim is to enable users to easily find the tweets of other users discussing similar topics, or inversely to make one's tweets more discoverable by others, and to see real time trends on the platform.
Hence, there can be completely different motivations behind writing a hashtag: to actually tag a tweet with one or more topics, to promote the tweet, or simply follow a trend.
Thus, the content of hashtags can deviate significantly from normal speech \cite{PageLinguisticsSelfbranding2012}.
It is therefore safer to discard hashtags entirely, which is no issue as long as we can collect enough textual content without them anyway.
We actually made some measurements in our tweets' database to see if that was the case.
We took several random samples of a million tweets each, stripped them of URLs and mentions, and then computed the ratio of characters within a hashtag compared to the total number of characters left in those tweets.
This proportion was found to be consistently below \SI{5}{\percent}.
We thus consider the precaution of stripping hashtags off of tweets worth taking.
After this cleaning step, for what follows we then keep only the tweets still containing at least four words.
The next crucial step is to infer the language the tweets are written in.
To do so, we leverage a trained neural network model for language identification: the Compact Language Detector \cite{SalcianuCompactLanguage2023a}, whose output is a language prediction along with the confidence of the model.
Subsequently, we only keep tweets detected as having been written in English with a confidence above \SI{90}{\percent}.
We subsequently pass these remaining tweets through LanguageTool to compute the absolute frequencies of non-standard features.
We then compute the frequency of non-standard features per word written by each user.
As the activity of Twitter users was previously found to follow a log-normal distribution spanning almost four orders of magnitude \cite{MocanuTwitterBabel2013}, we compute this frequency at the single user level to be more representative of the general population, and not only of the few very active users.
In any case, we remove inactive users and keep only ones who have written at least $100$ words.
Then, at the cell level, we compute the average of the individual relative frequencies for all residents.
To exclude cells with too little statistics, we only kept cells with at least $15$ residents left after applying the previous filter.
This leaves us with \SI{131402}{} users spread across \SI{4879}{} cells.
For reasons we will outline in the next section, in our study we concentrate on eight metropolitan areas around London, Manchester, Birmingham, Liverpool, Leeds, Bristol, Newcastle upon Tyne, and Sheffield.
In Table~S3 we give the precise definitions for these metropolitan areas. % \cref{SItab:uk_metropolitan_areas}

\begin{figure}
\centering
  \includegraphics[width=\textwidth]{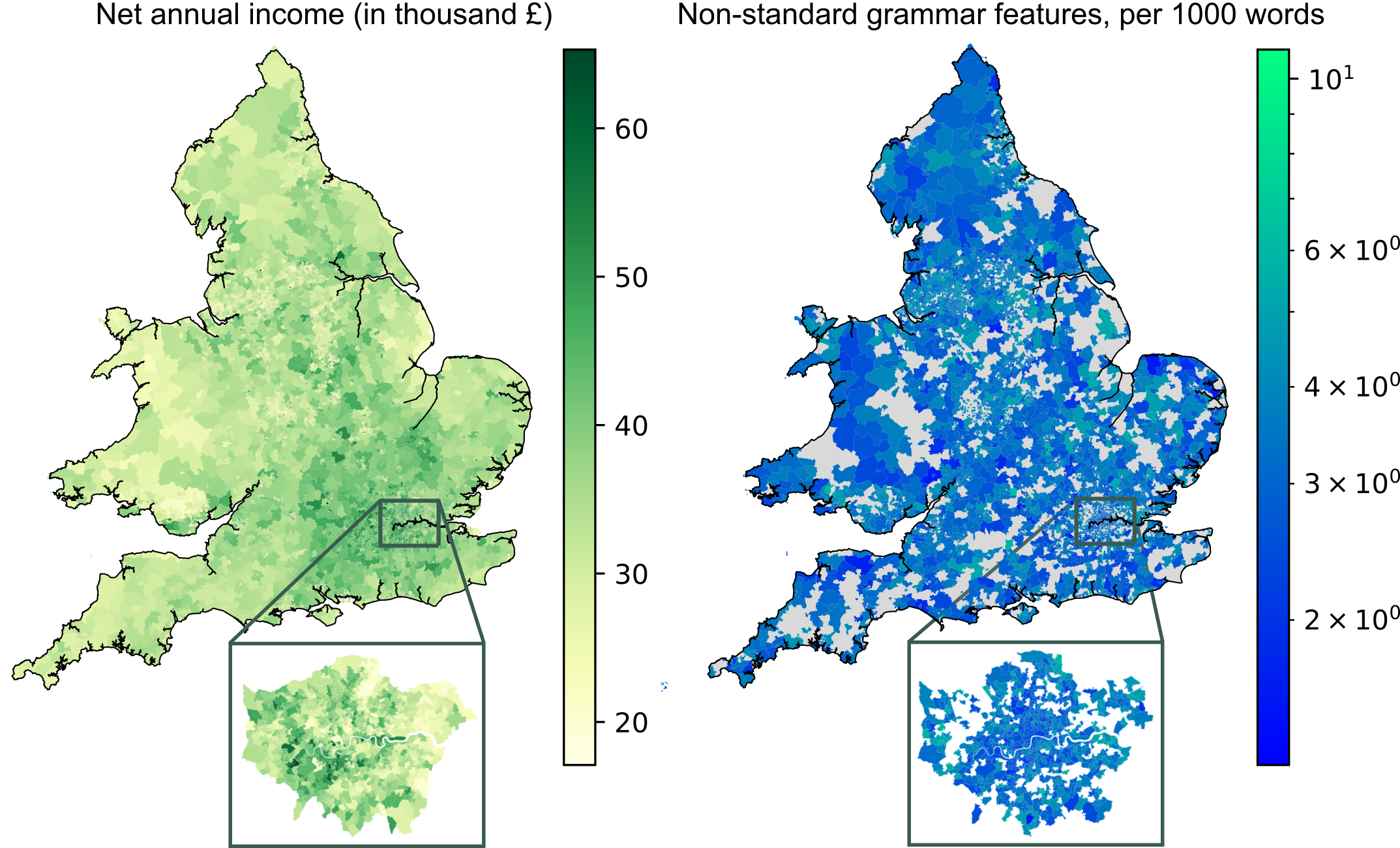}
  \caption{\textbf{Side-by-side maps of socioeconomic and linguistic variables.} They present the average net annual income given by the census (left), and the average frequency of non-standard features (right) measured from our Twitter dataset in the MSOAs of England and Wales. For the latter, we only colour cells with sufficient Twitter population, following the residence attribution, and a logarithmic scale is used to better visualise geographical variations. As an example, for both maps we provide a zoom-in on London showing the variation of these two variables in the city.}
  \label{fig:maps_income_vs_grammar_mistakes}
\end{figure}

At the end of this pipeline, in every remaining cell, our analysis yields an estimate of the income of its residents, which serves as a proxy for their SES, and of the frequency of the relative frequencies of their non-standard forms, which indicates how much they tend to deviate from a standard usage of English.
These two features are mapped side-by-side in \cref{fig:maps_income_vs_grammar_mistakes}. The results from our data processing are summarised in Table~S4, and are available in the form of aggregated data deposited in a public repository~\cite{LoufFrequencyDetected2023}. %\cref{SItab:dataset_summary}

\section{Empirical results}

\subsection{Linguistic varieties and socioeconomic status}

\begin{figure}
\centering
  \includegraphics[width=\textwidth]{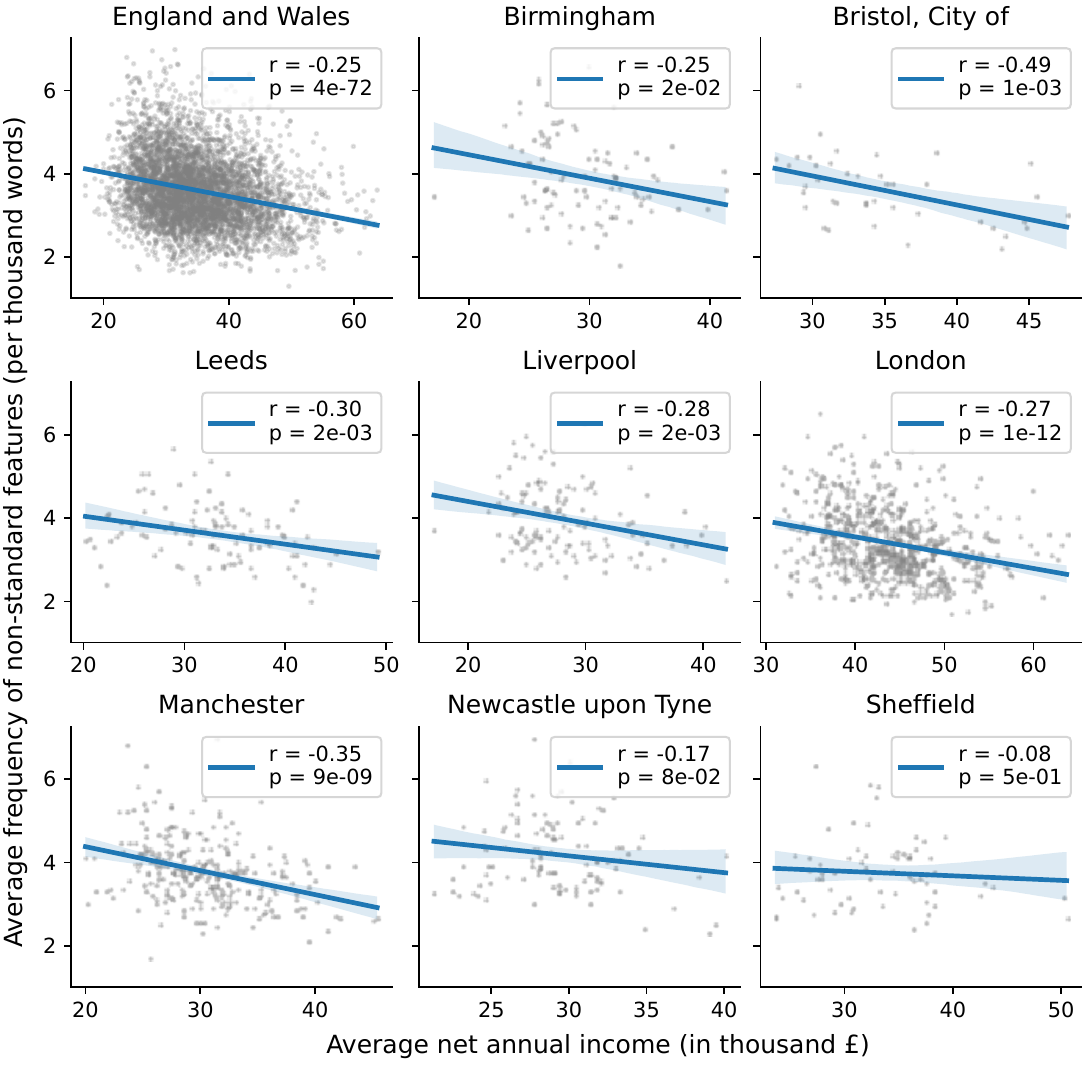}
  \caption{\textbf{Correlation between frequency of non-standard grammar features and net income.} Each data point corresponds to a different MSOA in England and Wales, and then in the $8$ metropolitan areas under consideration. Blue lines indicate the result of a linear regression, with the corresponding Pearson r and p-value given in each legend. The areas coloured in light blue indicate the \SI{95}{\percent} confidence interval for the regression estimates.}
  \label{fig:ses_x_grammar_corrs_all_cities}
\end{figure}

We recall that, according to earlier studies, the income of people and the frequency at which their language departs from standard grammar rules should be correlated negatively, thus indicating that higher income people would use non-standard features less frequently.
This was verified in very different contexts: some rather casual conversations carried out in a professional context \cite{LabovSocialStratification1966}, in formal interviews, reading and informal production \cite{TrudgillLinguisticChange1974}, and also in a microblogging context \cite{AbitbolSocioeconomicDependencies2018}, which mixes formal and informal speech in ways which are hard to disentangle.
This relation is indeed verified here in a context closest to the latter, as we measure a low but significant negative correlation (Pearson $r=-0.25$) when considering the whole observed population in England and Wales (as shown in \cref{fig:ses_x_grammar_corrs_all_cities}).
We would also expect similar correlations when focusing on urban populations.
Since most of our Twitter users are concentrated in urban areas, we expect more linguistic and socioeconomic heterogeneity in these places than in rural areas~\cite{MisloveUnderstandingDemographics2011}.
We therefore consider the eight largest metropolitan areas in England, listed earlier, as they have a sufficient number of cells remaining after the thresholding applied on the number of residents.
We summarise the observations in \cref{fig:ses_x_grammar_corrs_all_cities} together with the Pearson correlation coefficients and p-values in the inset.
Our results almost all reflect the expected pattern which suggests that speakers of high income favour the use of the standard variety and thus deviate less from the linguistic norm. However, quite remarkably, we find conspicuous differences among the cities, even though the ranges covered by their SES distributions are roughly similar. For example, in the case of Sheffield we find a very weak Pearson correlation $r = -0.08$, while in other cities much stronger correlations emerged, with correlation values up to $r = -0.49$ in Bristol. The largest cities like London and Birmingham, which host the most diverse populations, depict relatively strong linguistic correlations, with coefficients $r = -0.27$ and $r = -0.25$ respectively.
The standard deviations of the reported Pearson r values can be estimated, and they are shown in \cref{fig:cities_assort_vs_grammar_and_SES_corr}B.
Since we assess the correlation between two variables sampled in $N_c$ cells, the standard deviation can be estimated by
\begin{equation}
  \label{eq:pearsonr_std}
  \sigma_r = \frac{1 - r^2}{\sqrt{N_c - 3}},
\end{equation}
following the approximation from \cite{BonettMetaanalyticInterval2008}, shown to be unbiased even for small sample sizes in \cite{GnambsBriefNote2023}.

In multilingual societies, the interaction between standard and non-standard varieties is complex.
To analyze the possible influence of multilinguals in our corpus, we perform an analysis similar to the one described above.
Our methodology to obtain the proportions of multilinguals in every cell is outlined in section S5.
As shown in Table S5, it does not correlate significantly with the frequency of non-standard features.
This enables us to safely neglect the effect of multilingualism in our analysis.

\subsection{Assortative mixing and language variation}
\label{sec:assort_vs_ling}
To understand better the origin of the differences between the observed metropolitan areas, we study the mobility mixing patterns of their population. Following the areas visited by individuals in their daily mobility, we can observe how much people from different socioeconomic classes may meet and mix with each other in a given urban environment. Arguably, this may affect the language variety they adopt, assuming that segregated groups may influence less each other and adopt different varieties, while well-mixed populations may speak a similar language. To quantify mobility mixing in cities we measure their mobility assortativity~\cite{HilmanSocioeconomicBiases2022} in terms of movements between locations of different SES.
%To find out what could make these cities so different in that regard, we measure the assortativity in the mobility patterns of their residents \cite{HilmanSocioeconomicBiases2022}.
%We thus determine how likely people from different socio-economic classes are to interact with each other.
To establish this measure, for the same population of users we first take the inferred SES of each one of them and segment them into $n_{\sigma}$ equally populated SE classes.
Since the assigned SES indicator, i.e. average income, is the same for every user living in the same MSOA, they will all necessarily be assigned to the same class. Considering the cells in a given metropolitan area, we rank them by increasing average net income. We get their population from the census, denoted $N_c$ for each cell $c$ in the following. Denoting $I_c$ the set of cells with an average income
lower than or equal to $c$'s, we determine the SE class of $c$ as:
\begin{equation}
  \sigma_c = n_{\sigma} \left\lceil \frac{\sum_{c' \in I_c} N_{c'}}{\sum_{c'} N_{c'}} \right\rceil,
\end{equation}
with $\left\lceil \cdot \right\rceil$ representing the ceiling function and $n_{\sigma}$ the number of classes we wish to define. $\sigma_c$ takes integer values between $1$ and $n_{\sigma}$, both included, the former corresponding to the cells of lowest income, and the latter to the ones of highest income.

\begin{figure}
  \centering
  \includegraphics[width=\textwidth]{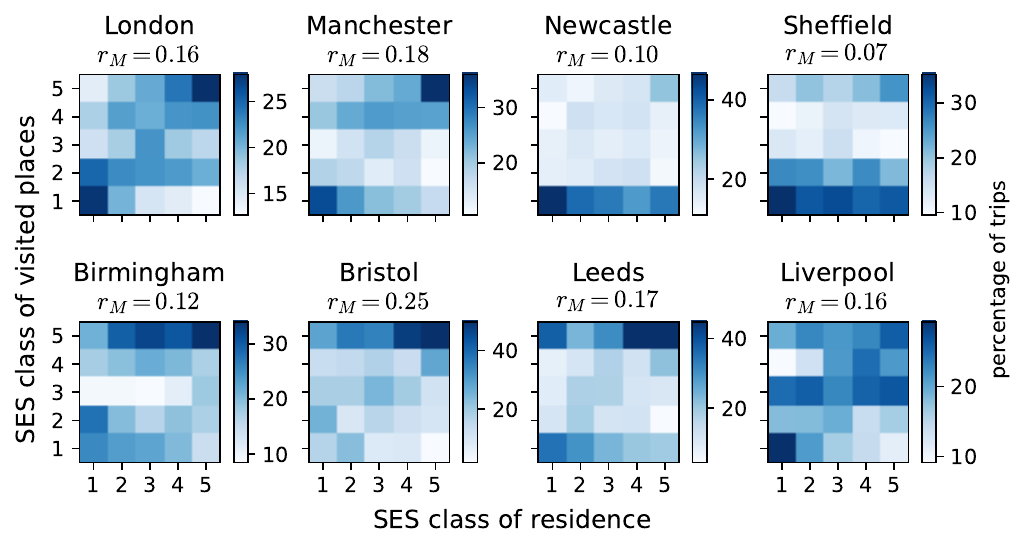}
  \caption{\textbf{The $M_{k, l}$ mobility mixing matrices stratified by SES in eight metropolitan areas of the UK.} Each matrix is column-wise normalised and depicts the probability for a user who resides in a location of class $l$ (x-axis) to make trips to a place of class $k$ (y-axis). For the definition of $M_{k, l}$, see \cref{eq:class_visit_proba}. For each metropolitan area, we give the $r_M$ assortativity coefficient measured from the corresponding mobility mixing matrix, as defined in \cref{eq:def_mobility_r}.}
    \label{fig:cities_SES_mobility}
\end{figure}

Next, we record $t_{u, c}$, the proportion of trips made by a user $u$ to cell $c$, by counting the number of geotagged tweets posted by user $u$ in every cell $c$.
The same set of geotagged tweets as the one used to infer the users' residence is used to compute $t_{u, c}$.
This allows us to introduce the probability for an individual to visit a cell of class $k$ knowing that they reside in a cell of class $l$:
\begin{equation}
  \label{eq:class_visit_proba}
  M_{k, l} = \frac{
      \sum_{c^* \in S_l} \sum_{u \in c^*} \sum_{c \in S_k} t^*_{u, c}
    }{
      \sum_{c^* \in S_l} \sum_{u \in c^*} \sum_{k' = 1}^{n_{\sigma}} \sum_{c \in S_{k'}} t^*_{u, c}
    },
\end{equation}
where $t^*_{u, c}$ is $t_{u, c}$ but set to zero for $c = c_u^*$, $c_u^*$ being the residence cell of user $u$, and renormalised so that $\sum_c t^*_{u, c} = 1$ for every user.
The $M_{k, l}$ form a column-wise normalised matrix, meaning $\sum_k M_{k, l} = 1$.
We here use the proportions $t^*_{u, c}$ instead of raw counts so as to make every user of a class contribute equally to $M_{k, l}$, thus accounting for the wide range of activity distributions.
Note that while computing this matrix we disregard stays in the cell of residence of the actual user. This way we remove local spatial effects that could induce spurious observations of strong mobility assortativity. The matrices computed for $n_\sigma = 5$ classes in the eight metropolitan areas are shown in \cref{fig:cities_SES_mobility} and depict some interesting patterns. First of all, it is evident that in many metropolitan areas like London, Manchester or Liverpool, the most segregated areas are at the poorest and richest locations. At the same time, in these cities the rest of the population is also segregated in terms of mobility, as indicated by the emergent diagonal component. This signals strong assortative mixing patterns in these cities, where people most likely mix with people of their own SE class, while being less likely to meet with dissimilar others. At the same time, in some cities like Newcastle, Sheffield, Birmingham or Bristol, mobility patterns are strongly biased towards one class (typically the richest or the poorest one). Such biases may appear due to geographic constraints (e.g. by a river separating the city) or due to urban design (e.g. having a shopping mall in one specific neighbourhood). In any case, none of these matrices is random, and each of them shows some specific mixing patterns that may indicate different levels of assortative mixing and mobility segregation.

\begin{figure}
\centering
\includegraphics[width=\textwidth]{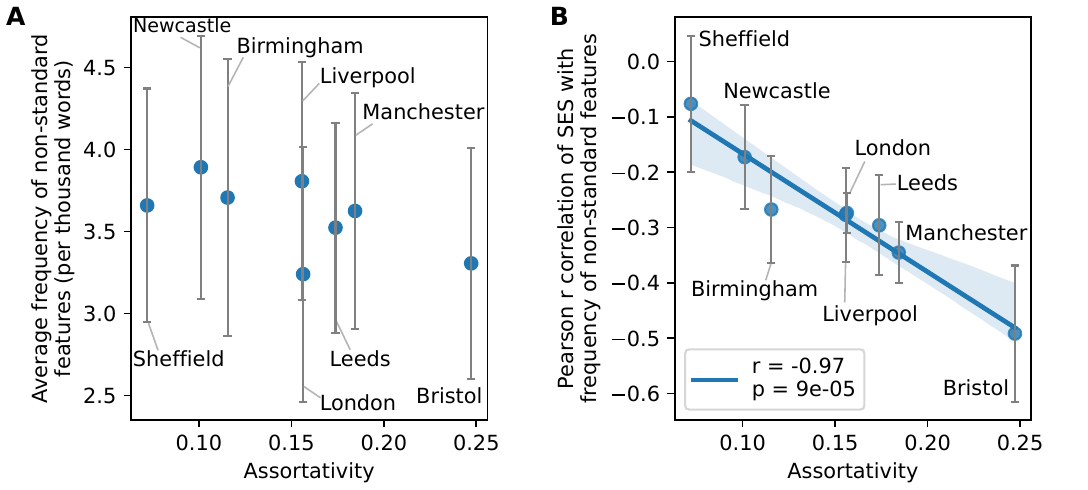}
  \caption{\textbf{The influence of assortativity on non-standard language use.} (\textbf{A}) At the level of metropolitan areas, the frequency of non-standard grammatical features does not correlate with assortativity. Indeed, the standard deviations shown with error bars are all larger than the differences between the city averages. (\textbf{B}) The dependence on SES of non-standard usage clearly decreases as assortativity increases. This means that the more SE classes mix with one another, the less the usage of the standard form of individuals will depend on their own SES. Error bars represent the standard deviations of the Pearson r values, estimated via \cref{eq:pearsonr_std}. The blue line indicates the result of a linear regression, with the corresponding Pearson r and p-value given in the legend. The area coloured in light blue indicates the \SI{95}{\percent} confidence interval for the regression estimates.}
  \label{fig:cities_assort_vs_grammar_and_SES_corr}
\end{figure}

The assortativity level present in these mobility mixing matrices can be quantified by the network assortativity coefficient $r_M$, which is defined as the Pearson correlation between values of the rows and columns of the matrix:
\begin{equation}
  \label{eq:def_mobility_r}
  r_M = \frac{
    N_M \sum_{k, l} k l M_{k, l}
      - \sum_{k, l} k M_{k, l} \cdot \sum_{k, l} l M_{k, l}
    }{
      \sqrt{N_M \sum_{k, l} k^2 M_{k, l} - \left( \sum_{k, l} k M_{k, l} \right)^2}
      \cdot \sqrt{N_M \sum_{k, l} l^2 M_{k, l} - \left( \sum_{k, l} l M_{k, l} \right)^2}
    },
\end{equation}
with $N_M = \sum_{k, l} M_{k, l}$, which is equal to the number of classes $n_\sigma$, by definition.
This coefficient takes values between $-1 \leq r_M \leq 1$. It is negative if people prefer to mix with dissimilar others, it is zero if the mixing is completely random without any bias, and it is positive in case of assortative biases, when people tend to mix with others of similar SES.
The $r_M$ assortativity coefficients measured from the corresponding matrices of the studied metropolitan areas are shown in the inset of each matrix in \cref{fig:cities_SES_mobility}.
These are computed for $n_\sigma = 5$ SE classes.
We present in Fig.~S2 the $r_M$ values for $n_\sigma \in \{3, 5, 10\}$, which show the robustness of $r_M$ under variation of $n_\sigma$ when the latter is high enough.
At the same time, they exhibit unexpected correlations when plotted against linguistic measures. % \cref{SIfig:assortativity_vs_nr_classes}
As shown in \cref{fig:cities_assort_vs_grammar_and_SES_corr}A, mobility mixing assortativity does not show any significant correlation with the average frequency of non-standard features measured in the different cities.
More strikingly, in \cref{fig:cities_assort_vs_grammar_and_SES_corr}B, we find a negative correlation between the mobility mixing assortativity and the correlation coefficient we measured earlier (see \cref{fig:ses_x_grammar_corrs_all_cities}) between SES and the average frequency of non-standard features.
This indicates that the stronger the mobility mixing we observe in a population, the less the propensity to deviate from standard rules is determined by the SES of origin.
Therefore, the use of standard English is not only determined by the SE class, but also importantly by the degree of mixing of the populations living in a metropolitan area.

\section{A model for dialect adoption}

\subsection{Definition}
Having identified the importance of social mixing, our next aim is to understand the mechanisms behind this observation with a simple quantitative model. The model should account for three main effects:
\begin{enumerate}[(i)]

  \item One of the two varieties of the language may be more prestigious than the other. This is for example the case of the standard form: it is taught at schools and spoken by mainstream media and public institutions~\cite{DavilaInevitabilityStandard2016,MilroySocialNetwork1992,MilroyIdeologyStandard2006}. This corresponds to what is known as \emph{overt prestige}.

  \item Even though one variety is less overtly prestigious, part of the population may still have a positive language attitude that makes them prefer using it. For instance, members of lower SES may prefer slang forms because they might provide them a sense of group identity. This is known as \emph{covert prestige}~\cite{LabovSocialStratification1966,TrudgillSocialDifferentiation1974,ChambersSocialDifferentiation2004}.

  \item We previously observed very different mixing patterns in various English metropolitan areas. Indeed, mobility can be very heterogeneous, so it should be possible to plug in any mobility data into the model to be able to understand how different mixing may affect the dynamics of the linguistic varieties.
\end{enumerate}

Relevant to our modelling challenge, there are already models for cultural transmission, like the seminal Axelrod model \cite{AxelrodDisseminationCulture1997,CastellanoNonequilibriumPhase2000,KlemmGlobalCulture2003,KlemmNonequilibriumTransitions2003,BattistonLayeredSocial2017}, that could be akin to our desired model for language variety adoption. However, in these models no dependence has been taken into account on an agent's intrinsic attribute or group identity, thus missing the effect of (ii) in their models. Other works have tried to model the diffusion of dialect features \cite{BurridgeSpatialEvolution2017,BurridgeInferringDrivers2021} or of whole languages \cite{KandlerLanguageShift2010,IsernLanguageExtinction2014}, which could also be considered similar to the dynamics we wish to understand here. But these models consider spatial diffusion with uniform use in areas, which work remarkably well to delimit dialectal regions, but are poorly adapted to fine-grained variations such as the ones we observe within metropolitan areas.

\begin{figure}
\centering
  \includegraphics{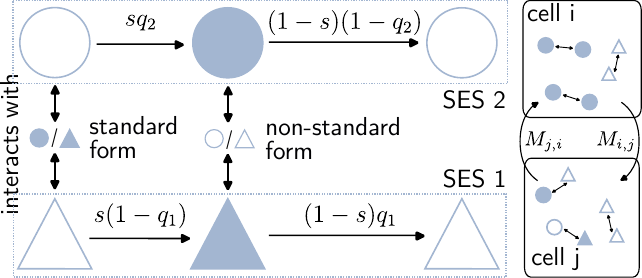}
  \caption{\textbf{Schematic summary of model definition.} The model features two SE classes, represented by circles and triangles. Agents in each class can speak one of two varieties of their language: the standard form (filled shapes) and the non-standard (empty shapes). Each agent has a cell of residence, and a mobility matrix $M_{i, j}$ that defines the probabilities for a resident of
  $i$ to be at different cells at each time step. Once an agent is moved, another agent is picked randomly to interact with. If they interact with an agent using the other variety, they adopt this variety with a probability that depends on their SE class
  of origin. For instance, when a triangle (SES 1) speaking non-standard interacts with another agent from any class speaking standard, they have a probability equal to $s (1 - q_1)$ to adopt the standard form. Possible transitions are presented as arrows between circles and triangles.}
  \label{fig:ses_ling_model_diagram}
\end{figure}

With these considerations in mind, we propose an agent-based model (ABM) as follows (see \cref{fig:ses_ling_model_diagram} for the sketch of our model).
Our model considers agents who are assigned as residents to a cell, which belongs to one of two SE classes. Agents have a linguistic state, which can either be that they use the standard variety, or that they use a non-standard dialect. The overt prestige is operationalised with the variable $s$, which can take values from the unit interval. Note that in our case, to comply with effect (i) that we wish to model, we always set $s>0.5$, thus introducing a higher prestige for the standard form. To model effect (ii), each SE class has a preference towards one form: class 1 is biased towards a non-standard form (denoted 1) with a factor $q_1$, and inversely class 2 is attached to the standard one (denoted 2) with a factor $q_2$. These factors are also constrained between $0$ and $1$, and when they take a value above $0.5$, they represent a bias towards the respective form. For instance, when an agent of class 1 speaking non-standard interacts with another agent speaking standard, they have a probability $s (1 - q_1)$ to start using the standard form as well.
For definiteness, here and in the following, we use the term ``non-standard form'' to aggregate together all varieties which are not the standard one.
We therefore use a simplifying dichotomy between standard and non-standard to focus on the propensity to deviate from the former, as we did previously in the empirical analysis.

Beyond the language adoption dynamics, we allow agents to move to different cells $j$, according to the probabilities $M_{i,j}$, conditional to their cell of residence $i$. This mechanism controls the mixing of the two populations discussed in (iii).

\subsection{Model analysis}
Our model can be simulated for any number of cells and arrangements of the populations of different SE classes. However, here we will first consider a simple case in order to present succinct equations that lend themselves to interpretation and may lead to a treatable mathematical description. We will consider only two cells, with all individuals of class 1 residing in cell 1, and all individuals of class 2 residing in cell 2. We assign by $p_1$ the proportion of individuals of class 1 speaking non-standard (variety 1), and by $p_2$ the proportion of individuals of class 2 speaking standard forms (variety 2). Individuals have to speak either 1 or 2, which implies that, for instance, a proportion $(1 - p_1)$ of individuals of class 1 speak the variety 2. These two variables therefore summarise the linguistic state of the system.
Regarding the mobility, we make the assumption that people of the two classes have the same probability $M$ to move away from their residence cell. Note that this parameter describes how much the classes mix, as the Pearson $r$ of the corresponding mobility
matrix as defined in \cref{eq:def_mobility_r} satisfies $r_M = 1 - 2 M$.
We also assume that the two SE classes have the same population, and that speakers' interactions are all-to-all..
In other words, we adopt a mean-field approximation, which is a good approximation when considering large populations.
After following the steps detailed in section S6, we can then describe our model with the following system of coupled differential equations: %  \cref{SIsec:model_analytic}
\begin{equation}
  \label{eq:ses_ling_time_evol_eq_mob}
  \left\{
  \begin{aligned}
      \dv{p_1}{t}
          &=
            2 M (1 - M) (1 - p_1 - p_2) [q_1 (1 - s) - p_1 (q_1 - s)]+
            p_1 (1 - p_1) (q_1 - s)
      \\[1ex]
      \dv{p_2}{t}
          &=
            2 M (1 - M) (1 - p_1 - p_2) [q_2 s - p_2 (s + q_2 - 1)]+
            p_2 (1 - p_2) (s + q_2 - 1)
  \end{aligned}
  \right.
\end{equation}

Interestingly, each equation features a first term linked to the group mobility, maximum for $M = \frac{1}{2}$, which describes maximum mixing. This term disappears when $p_1 = 1 - p_2$ and leads to convergence in the usage of language varieties between SE classes. Indeed, in the first equation, if $q_1 \leq s$, the term in square brackets is clearly strictly positive. If $q_1 > s$, we have
\begin{equation}
  q_1 (1 - s) - p_1 (q_1 - s) > q_1 (1 - s) - (q_1 - s) = s (1 - q_1) > 0.
\end{equation}
As a consequence, the sign of the mobility term follows the one of $(1 - p_1 - p_2)$. It is therefore negative if $p_1 > 1 - p_2$ and positive otherwise, thus pushing $p_1$ towards $1 - p_2$. Similarly, in the second equation, since $q_2 s - q_2 - s > -1$, the mobility term pushes $p_2$ towards $1 - p_1$.
The second term of each evolution equation represents ``self-growth'', independent of
mobility and maximum for $p_k = \frac{1}{2}$. This term thus brings a given class $k$ towards homogeneity. For $k = 1$ for instance, it does it either towards $p_1 = 0$ for $q_1 < s$, or towards $p_1 = 1$ for
$q_1 > s$. When there is no mixing, for $M = 0$ or $M = 1$, as expected the two populations become completely independent, and the only stable fixed points of the system correspond to homogeneous populations in terms of the language variety they use. The emerging dominant variety depends on the sign of $(q_1 - s)$ and $(s + q_2 - 1)$ for classes 1 and 2, respectively.
To summarise, two opposing effects can be identified in the dynamics of the model in this simple setting.
\begin{itemize}
  \item The more individuals of different classes mix through their mobility patterns, the smaller the differences in the usage of language varieties between them. This aligns with our observation from \cref{fig:cities_assort_vs_grammar_and_SES_corr}B.
  \item The stronger the bias of a class for its preferred variety, the more homogeneous the variety adoption within this class.
\end{itemize}

\begin{figure}
\centering
  \includegraphics{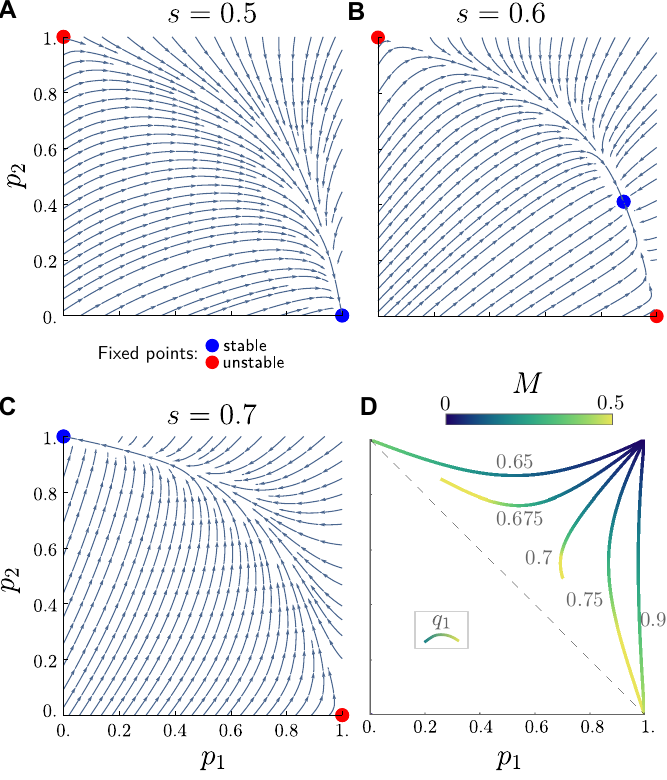}
  \caption{\textbf{Stability analysis of the model dynamics.} We analyse the mean-field description of the model, given in \cref{eq:ses_ling_time_evol_eq_mob}, with $q_2=0.5$.
  The analysis indicates what values of $p_1$, the proportion of individuals of class 1 speaking non-standard (variety 1), and $p_2$, the proportion of individuals of class 2 speaking standard (variety 2), are reached by the model as it converges.
  Stream plots show the fixed points of the model for a mobility mixing $M = 0.2$, a $q_1 = 0.7$ and (\textbf{A}) $s = 0.5$, (\textbf{B}) $s = 0.6$ and (\textbf{C}) $s = 0.7$.
  The points $(1, 0)$ and $(0, 1)$ are shown to consistently appear as stable (blue) or unstable (red) fixed points.
  The first is stable for $s$ significantly lower than $q_1$ (as in \textbf{A}) and the second for $s$ close to or higher than $q_1$ (as in \textbf{C}).
  A third fixed point featuring coexistence of the two varieties may appear for intermediate values of $s$ (as in \textbf{B}).
  (\textbf{D}) After setting $s = 0.6$, we show trajectories of the stable fixed points for different values of $q_1$, corresponding to the coexistence when varying $M$ from $0$ to $0.5$. This shows how increasing the mobility mixing $M$ moves the stable fixed point of coexistence toward the dashed line corresponding to $p_2 = 1 - p_1$, and towards the extinction of variety 1 or 2, for respectively low and high values of $q_1$.}
  \label{fig:mean_field_analytic_coex_pos}
\end{figure}

To characterise the fixed points of this system of equations, we also perform a stability analysis. There are two fixed points that can be trivially found from \cref{eq:ses_ling_time_evol_eq_mob}: $(p_1, p_2) = (0, 1)$ and $(1,0)$. Other fixed points are located using symbolic computations. Then, to determine whether these fixed points are stable states of convergence of the system, we compute the eigenvalues of the Jacobian of the system evaluated at its fixed points. The stable fixed points are those whose corresponding eigenvalues have a strictly negative real part.
We thus find that the model can not only converge to a state featuring the extinction of one of the two varieties, but also that both may coexist. To illustrate this point, for $M
= 0.2$, $q_2= 0.5$, $q_1 = 0.7$, and $s \in \{ 0.5, 0.6, 0.7\}$, we show in
\cref{fig:mean_field_analytic_coex_pos}A-B-C the position of the identified fixed points of the model and whether they are stable. These stream plots also demonstrate the dynamics leading to convergence.

When the covert prestige of non-standard forms parametrised by $q_1$ is significantly higher than the overt prestige $s$ of the standard variety (as in panel (a)) the stable fixed point corresponds to the dominance of variety 1. At the same time, $q_1=s$ implies a stable dominance of variety 2 (as shown in panel (c)). Coexistence of the two varieties is possible for intermediate values of $s < q_1$, like for $0.6$ (as shown in panel (b))
As proven in Section~S6.5, the condition $s < q_1$ is actually a strong requirement for the existence of a fixed point corresponding to coexistence. % \cref{SIsec:coex_sol_ses_model}
In \cref{fig:mean_field_analytic_coex_pos}D, we show how this stable fixed point moves in the $(p_1, p_2)$ space when varying $M$, for $s = 0.6$ and different values of $q_1$. As shown earlier, increasing $M$ pushes the system towards $p_2 = 1 - p_1$, and whether it is biased towards variety 1 or 2 depends on the difference between $q_1$ and $s$. This feature is very much in line with the observation we made in
\cref{fig:cities_assort_vs_grammar_and_SES_corr}: the less assortative the mixing -- that is the closer $r_M$ here is to $0$ -- the less the use of the standard variety depends on SES -- which in our case means that $(1 - p_1)$ and $p_2$ will get closer to one another.

\begin{figure}
  \centering
  \includegraphics{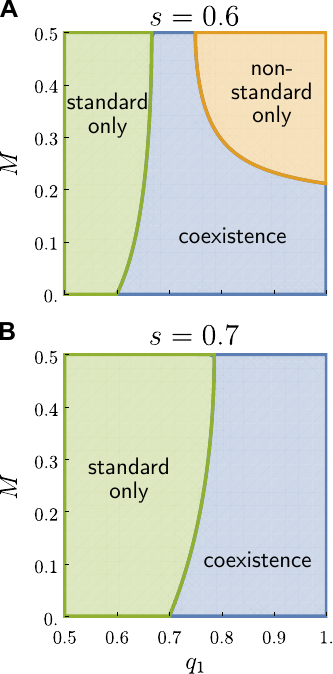}
  \caption{\textbf{Parameter space of the model in a simple setting split by the kind of stable fixed points.} We show the solutions for the mean-field model defined in
  \cref{eq:ses_ling_time_evol_eq_mob}, with $q_2$ set at the neutral value of $0.5$. We depict the regions of the $(q_1, M)$ parameter space where different kinds of stable solutions emerge, with (\textbf{A}) $s = 0.6$ and (\textbf{B}) $s = 0.7$. Regions correspond to three possible stable fixed points: the extinction of variety 2 and dominance of 1 labelled with ``non-standard only'', inversely the dominance of 2 labelled with ``standard only'', and the coexistence of both varieties labelled with ``coexistence''. Examples for each of these are respectively shown in \cref{fig:mean_field_analytic_coex_pos}A-C-B. Stable coexistence of varieties 1 and 2 is favoured for rather low mobility and a preference of class 1 for variety 1 $q_1$ higher than the overt prestige $s$.}
  \label{fig:mean_field_analytic_q-M_space}
\end{figure}

In \cref{fig:mean_field_analytic_q-M_space} we present the regions of the $(q_1, M)$ parameter space where the aforementioned stable solutions appear. For an overt prestige favouring variety 2 ($s > 0.5$), but with a value still low enough (as in panel (a)), three distinct domains appear in the parameter space: two corresponding to dominance of one variety and extinction of the other, and a third corresponding to coexistence of the two varieties. At the same time, higher values of $s$ prohibit stable solutions associated to the dominance of 1, as shown in panel (b). Coexistence is facilitated by less mixing between the different classes, yet for a given range of values of $q_1$, no matter how much populations mix, the two varieties can still coexist. This is a noteworthy result, because while segregation leads to the conservation of varieties, we find that coexistence is still possible regardless of the social mixing.

\subsection{Data-driven simulation for metropolitan areas}
\begin{figure}
\centering
  \includegraphics{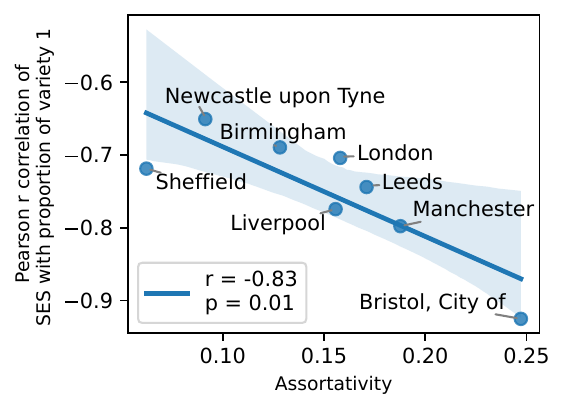}
  \caption{\textbf{Sample result from the ABM simulations in eight metropolitan areas.}
  Correlation between the average income and proportion of speakers choosing variety 1 at the stable state of convergence of the simulations for $s = 0.55$, $q_1 = 0.7$ and $q_2 = 0.5$ versus the cities' assortativity.
  This example was selected as an illustration, as it features one of the lowest correlations between x and y-axis values, which are shown for other values of $q_1$ and $s$ in \cref{fig:phase_space_ses_city_sims}.
}
  \label{fig:sample_city_sims_res}
\end{figure}

The rich phenomenology that the model exhibits calls for further investigation to see if it can reproduce the empirically-observed correlations between assortativity and language varieties, once initialised with real parameters. We run simulations in each of the eight metropolitan areas featured in \cref{fig:cities_SES_mobility,fig:cities_assort_vs_grammar_and_SES_corr}. We populate their MSOAs with as many agents as they have inhabitants according to the census, and attribute the average income also given by the census to all agents of each area. In these simulations, we consider five SE classes, so that we can compare results directly to the outputs of our data analysis, shown in \cref{fig:cities_assort_vs_grammar_and_SES_corr}.
To fully parameterise the model, we need three more linguistic parameters, namely the preferences for one of the two language varieties for each additional class. In order to limit the number of parameters to explore, we only keep the preferences $q_1$ and $q_2$, which set the preference of class 1 (the lowest SES) for variety 1 and of class 5 (the highest SES) for variety 2, respectively. Meanwhile, to attribute preferences for variety 1 to each remaining class, we use a linear interpolation between $q_1$ and $1 - q_2$. To be consistent with the phenomenon we are interested in, this interpolation is always a decreasing function of the class number, as we run simulations with $q_1 > 0.5$ and $q_2 \geq 0.5$. To avoid the uninteresting states of convergence featuring a hegemony of standard language, we only consider values of the overt prestige such that $s > 0.5$ and $q_1 > s$. Thus, the only feature left as unknown parameter for the simulations is the inter-cell mobility of the agents. To estimate this parameter, we leverage the mobility observed in our Twitter dataset, namely, the $t_{u, c}$ previously introduced, but this time including the travels of the users within their own cell of residence. We thus obtain a matrix giving for each cell the probability for their residents to be found in any cell at each simulation step. In summary, at most three random draws are performed at each step for each agent: i) to decide in which cell this agent will be interacting with others; ii) to choose the other agent they will interact with; and, if the two are using different varieties, iii) to decide whether they will switch, according to their class switch probability (similarly to what is depicted in \cref{fig:ses_ling_model_diagram} for just two classes). A sample result from our simulations is given in \cref{fig:sample_city_sims_res}.

\begin{figure}[hb]
\centering
  \includegraphics{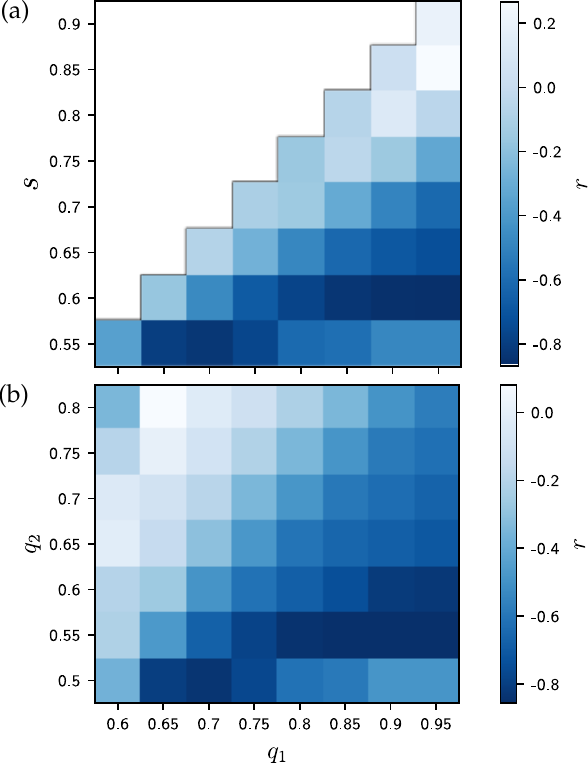}
  \caption[Phase-space exploration of the model.]{\textbf{Phase-space exploration of the model.} We show the correlations $r$ between the empirically-measured mobility assortativity and the correlation between income and proportion of usage of variety 1 (non-standard) measured at the end of simulations in all eight cities. These correlation values are shown for different values of $q_1$ and $s$ for $q_2 = 0.5$ in (a), and of $q_1$ and $q_2$ for $s = 0.55$ in (b).}
  \label{fig:phase_space_ses_city_sims}
\end{figure}

The original aim of the proposed model was to better understand how social mixing facilitates the interdependence between SES and the usage of standard language, that we have shown in \cref{fig:cities_assort_vs_grammar_and_SES_corr}. To synthesise an answer to this challenge, at the stable state reached by the simulations in each city, we measure in each cell the proportion of agents using the non-standard language variety 1. Similarly to the data analysis, we compute the correlation between the cells' average income and the proportion of agents using the non-standard form, and check it against the measured assortativity of the different cities (for model parameter values, see the caption of \cref{fig:sample_city_sims_res}).
By comparing \cref{fig:sample_city_sims_res} with the corresponding empirical results in \cref{fig:cities_assort_vs_grammar_and_SES_corr} we find striking similarities between the observed and modelled correlations, that verifies the modelled mechanisms to give a possible explanation for the observed phenomena. Note that due to the two different proxies for the usage of non-standard language, beyond the observed negative correlations, we do not expect the absolute values of the y-axes to match.
The parameter set that led to the result of \cref{fig:sample_city_sims_res} was not chosen arbitrarily.
Indeed, to find the best parametrization that most closely reproduce the empirical observations, we performed a grid search in the parameter space with an increment of $0.05$ in parameter values.
These are shown in \cref{fig:phase_space_ses_city_sims}.
As visible in \cref{fig:phase_space_ses_city_sims}, to avoid states of convergence that are irrelevant to us (featuring the extinction of variety 1), simulations were run only for values of $q_1$ strictly superior to $s$. % \cref{SIfig:phase_space_ses_city_sims}A
This panel depicts correlation values computed between the axes of plots similar to \cref{fig:sample_city_sims_res}, but for different parameter values.
The highest absolute correlation value in this plot corresponds to $s = 0.55$, $q_1 = 0.7$ and $q_2 = 0.5$.
The $r$ values are shown in \cref{fig:phase_space_ses_city_sims}A-B for $q_2 = 0.5$ and $s = 0.55$, respectively. % \cref{SIfig:phase_space_ses_city_sims}A-B
The two panels clearly demonstrate that the empirically-observed correlation pattern can be robustly reproduced for several parameter values, which lead to stable states featuring $r$ values relatively close to $-1$.
Consequently, our simple model with only three parameters is able to capture our empirical observations with notable fidelity.

\section{Discussion and conclusions}
Throughout this work, we have investigated the inter-dependence between SES and the usage of different linguistic varieties.
Focusing on the dichotomy between standard and non-standard English usage in eight UK cities, through a combination of Twitter and census data we found that the average frequency of non-standard features and average income are slightly correlated.
More interestingly, the more different SE classes mix together, the weaker this correlation, meaning the more similar the usage of English across classes.

We have subsequently introduced an ABM that proposes an explanation for this observation.
It features transition probabilities from using one variety to the other that depend on both a globally shared overt prestige of the standard, and a class-dependent covert prestige --- or preference --- for either the standard or the non-standard dialect.
We have here shown quantitatively that the positive language attitude of a group towards a linguistic feature can be crucial to sustain it, despite an unfavourable overt prestige.
The analytic framework we presented here has the virtue of enabling us to capture the role of social mixing between these different groups.
In accordance with our observations made with a large Twitter geotagged dataset, we have shown analytically that it tends to smooth out linguistic differences between SE classes.
Remarkably, this smoothing does not necessarily imply homogeneous language, but may also mean comparable prominence of the two linguistic varieties of the two groups.
These analytic findings are also supported by the results of agent-based simulations involving the populations of the eight metropolitan areas we studied empirically.
The simulations can yield the same relationship we obtained from the data, namely that the more social mixing there is across socioeconomic classes, the less the individuals' choice to use standard or non-standard language will depend on their class of origin.

This work provides a solid foundation for future works of the same vein.
It could first be extended to other countries where similar data could be obtained in sufficient amounts.
Further, since our data analysis is limited to a single data source, investigating the same issues with alternative sources can only refine our understanding of the phenomenon.
As we already pointed out in the introduction, we do believe we made reasonable assumptions about the biases inherent to our Twitter dataset, but verifying our results using another source could help validate our assumptions and prove the robustness of the results.
Indeed, it could be for instance that since our sample is on average relatively young, we uncovered relationships that do not hold uniformly across all age groups.
Further, an in-depth analysis of how the status of immigrant may impact the use of standard language is another aspect this study could not address.
It is a question that is both intriguing and deeply complex to approach, because being an immigrant can mean very different things, such as having a multilingual repertoire, a use of the English language that is mostly confined to formal contexts, a partial knowledge of the language, or the prior knowledge of a variety that is different from standard British English.
Also, Twitter is a particular social context, but individuals may choose to use a different language in other environments.
This potential change of behaviour is therefore not captured by our model, but it could be relevant to the global dynamics.
Observing the language production of individuals in different social contexts on a scale such as the one we presented here poses a great challenge, but it would definitely help further modelling endeavours and thus greatly contribute to our understanding of these linguistic phenomena.
Moreover, we defined SES from income only, while it is multifaceted, encompassing individuals' income, education, professional occupation or even accumulated wealth, to cite a few.
A more complete, but also complex, characterisation of SES in further works would greatly refine the analysis as well.
Besides, the model we propose, whose strength is to be able to reproduce a non-trivial observed pattern from very simple mechanisms, could also be extended to more precisely capture sociolinguistic phenomena.
In particular, we used a simple dichotomy between overt and covert prestige, that could be extended to account for the presence of several group-based norms that coexist in parallel.
This is for instance the case for groups of non-native speakers, who have more than one set of linguistic standards.
Besides, the model assumes the choice of a single linguistic variety by each speaker.
This necessarily obscures code-switching practices, which can themselves relate to socio-economic class, and would therefore need proper consideration.

Still, our findings have uncovered yet another effect of segregation in mobility patterns, that is how it can affect language use across socioeconomic groups.
This work thus echoes a trail of others which have shown the importance of tackling mobility segregation in urban areas \cite{NieuwenhuisDoesSegregation2020,MoroMobilityPatterns2021,NilforoshanHumanMobility2023,IyerMobilityTransit2023}.
It also sheds light on how central mobility mixing patterns can be for sociolinguistic studies, and thus shows that this variable should not be ignored.

\section*{List of abbreviations}
\begin{itemize}
  \item ABM: agent-based model
  \item MSOA: middle layer super output areas
  \item SE: socio-economic
  \item SES: socio-economic status
\end{itemize}

\section*{Declarations}

\subsection*{Ethics approval and consent to participate}
Not applicable.

\subsection*{Consent for publication}
Not applicable.

\subsection*{Data availability}
All aggregated data we have produced for this paper are deposited in an Open Science Framework repository \cite{LoufFrequencyDetected2023}, and can be downloaded from \url{https://osf.io/rsxud/}.
The boundaries of the MSOAs of England and Wales can be downloaded from the Open Geography portal of the Office for National Statistics: \url{https://geoportal.statistics.gov.uk/datasets/ons::middle-layer-super-output-areas-december-2011-boundaries-ew-bgc-v3-1/about}.
The estimates of the 2018 census for the net annual income of the MSOAs of England and Wales can be downloaded from \url{https://www.ons.gov.uk/employmentandlabourmarket/peopleinwork/earningsandworkinghours/datasets/smallareaincomeestimatesformiddlelayersuperoutputareasenglandandwales}, while their population can be obtained from \url{https://www.nomisweb.co.uk/census/2011/ks101uk}.

\subsection*{Code availability}
All code needed to reproduce the results in the paper is deposited in public Zenodo records: both for the Python code used to analyse data, run the simulations and produce all figures except \cref{fig:mean_field_analytic_coex_pos,fig:mean_field_analytic_q-M_space} \cite{LoufTLoufSesling2024}, and for the Mathematica code used to analyse our model in mean-field and produce the two aforementioned figures \cite{LoufTLoufSeslinganalytical2024}.

\subsection*{Competing interests}
The authors declare that they have no competing interests.

\subsection*{Funding}
This work was partially supported by the Spanish State Research Agency (MCIN\slash AEI\slash 10.13039\slash 501100011033) and FEDER (UE) under project APASOS (PID2021-122256NB-C21 and PID2021-122256NB-C22) and the María de Maeztu project CEX2021-001164-M and by the Comunitat Autonoma de les Illes Balears through the Direcció General de Política Universitària i Recerca with funds from the Tourist Stay Tax Law ITS 2017-006 (PDR2020/51). TL is grateful for the support from the EMOMAP CIVICA project. MK was supported by the CHIST-ERA project SAI: FWF I 5205-N; the SoBigData++ H2020-871042; the EPO and EMOMAP CIVICA projects and the DATAREDUX project, ANR19-CE46-0008.

\subsection*{Authors' contributions}
{\parindent0pt

Conceptualization: TL, JJR, DS, MK

Data curation: TL

Formal Analysis: TL

Funding acquisition: JJR, DS, MK

Methodology: TL, JJR, DS, MK

Software: TL

Supervision: JJR, DS, MK

Validation: TL, JJR, DS, MK

Visualization: TL, DS, MK

Writing -- original draft: TL

Writing -- review \& editing: TL, JJR, DS, MK
}

\subsection*{Acknowledgements}
Not applicable.

\bibliography{biblio}
% \printbibliography

\end{document}

\baselineskip24pt

\maketitle

\clearpage

\tableofcontents
\listoffigures
\listoftables

\clearpage
% are following sections actually useful?
\section{Description of the dataset}
\Cref{tab:dataset_summary} gives some summary statistics of our filtered Twitter dataset, obtained after going through all the pre-processing steps described in the main text. These show some variation from one metropolitan area to another, but, reassuringly, the user averages are quite consistent across the board.

\begin{figure}[p]
\centering
  \includegraphics[width=1\textwidth]{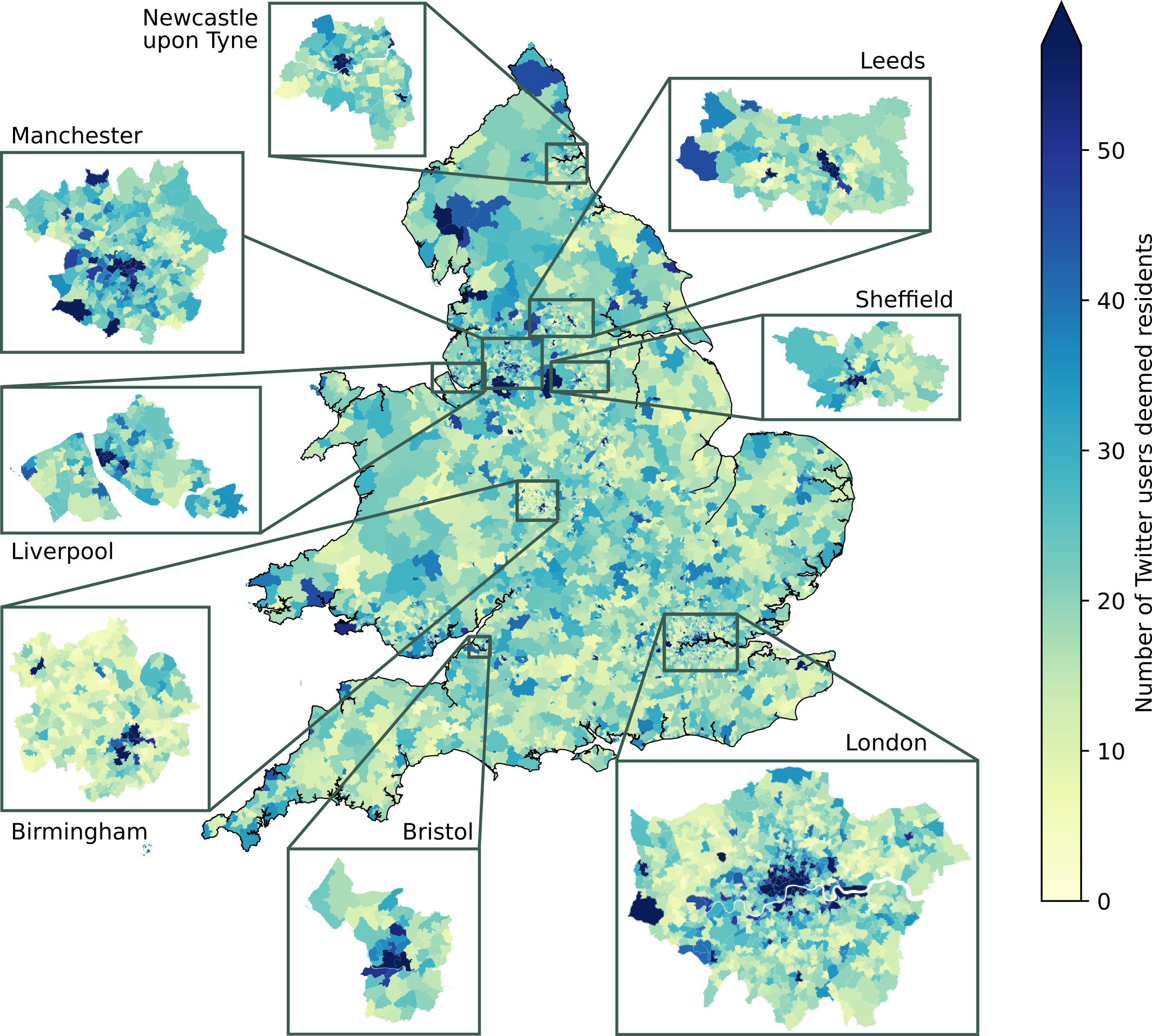}
  \caption[Twitter population map in England and Wales.]{\textbf{Twitter population map in England and Wales.} The users counted in each
  MSOA were deemed residents of the areas. A zoom-in on each of the eight metropolitan areas of the study shows the MSOAs selected in their definition, which are also given in \cref{tab:uk_metropolitan_areas}.}
  \label{fig:EW_twitter_pop}
\end{figure}

We also show a map presenting the population of each MSOA of England and Wales in \cref{fig:EW_twitter_pop}.

\section{Statistics for the categories of standard language rules}
\label{sec:cats_stats}
\Cref{tab:cats_global_stats} gives the number of matches for the rules of each category defined by LanguageTool on our filtered corpus, as well as the computed Pearson r correlation of their user-averaged frequencies with the average net income in the MSOAs of England and Wales. The very high number of typographical deviations is mostly due to the presence of extra whitespaces induced by our filtering of URLs, hashtags and mentions.

We subsequently focused on grammar features, which seem the most anti-correlated with net income. The top ten rules from that category are given in \cref{tab:top_grammar_mistakes}, to give an idea of the kind of rules that served us as a proxy to quantify deviations from the standard variety.

More information about LanguageTool's rules and categories can be found at \url{https://community.languagetool.org/rule/list}.

\section{Metropolitan areas' definition}
\Cref{tab:uk_metropolitan_areas} details explicitly the areas included in our definitions of the eight metropolitan areas studied throughout this work.
Clearly, all areas show a similar number of tweets, tokens and frequencies of non-standard features per user, which makes it possible to make comparisons between cities reliably.

\section{Assortativity's dependence on the number of classes}
\begin{figure}[p]
\centering
  \includegraphics{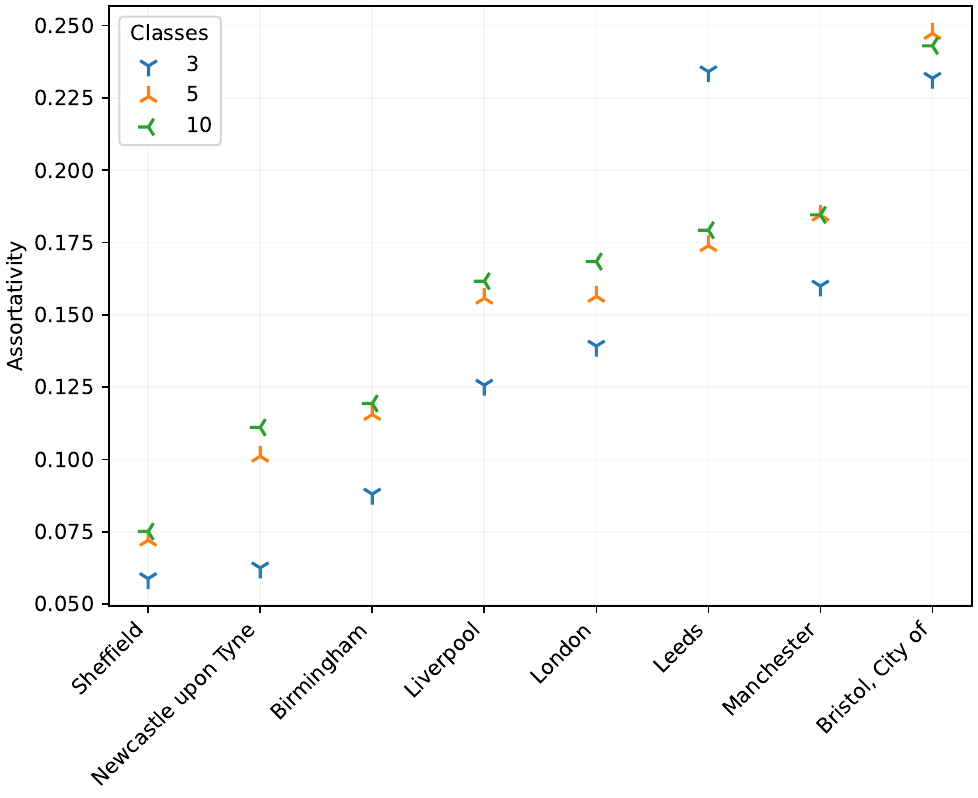}
  \caption[The influence of the number of classes on the computed assortativity.]{\textbf{The influence of the number of classes on the computed assortativity.}}
  \label{fig:assortativity_vs_nr_classes}
\end{figure}

In the main text, we give assortativity values in our eight metropolitan areas that were computed after defining five socioeconomic classes. We show in \cref{fig:assortativity_vs_nr_classes} the values of the assortativity for three and ten classes, which show the robustness of our measurement when the number of classes is not too small.

\section{Influence of multilingualism}
In order to uncover a potential impact of the potential multilingual backgrounds of our home-located users on our study, we performed the following analysis.
We gathered all the tweets from our identified residents and performed a language detection on each of them.
To do so, we first cleared them of hashtags, URLs and mentions, and then kept those tweets which retained at least 4 words.
All tweets which passed this threshold are then passed into Chromium's Compact Language Detector (CLD), which, for each tweet, gives us a language and the confidence of the algorithm.
We therefore only kept those with a confidence above $90\%$, to then count for each user the number of times they tweeted in different languages.
For each user, we only keep languages that either appear in at least $3$ tweets or $10\%$ of them, as some may occasionally use a translator or quote someone else in a language they cannot speak.
The results are presented in \cref{tab:multiling}.

\section{Analytic results for our model of variety adoption}
\label{sec:model_analytic}

\subsection{Notation}
Let us introduce the following notation:
\begin{itemize}
    \item $\Sigma$ the set of SE classes in a population:
        \begin{equation}
            \Sigma = \left\{ \sigma_k \mid k \in [1, n_{\sigma}] \right\},
        \end{equation}
    \item $C$ the set of cells of residence:
        \begin{equation}
             C = \left\{ c_i \mid i \in [1, n_C] \right\},
        \end{equation}
    \item $N_{c, \sigma}$ the number of residents of cell $c$ with class $\sigma$,
    \item $N_{c} \equiv \sum_{\sigma} N_{c, \sigma}$ the population of cell $c$,
    \item $N_{\sigma} \equiv \sum_c N_{c, \sigma}$ the population of class $\sigma$,
    \item $N \equiv \sum_{c, \sigma} N_{c, \sigma}$ the total population,
    \item $M_{i, j}$ the probability for a resident of $c_i$ to move to $c_j$.
\end{itemize}

\subsection{Assumptions}
Let there be only two cells: $n_C = 2$, and two SE classes: $n_{\sigma} = 2$, completely
separated, with the whole $\sigma_1$ population in $c_1$ and the whole $\sigma_2$ population in
$c_2$:
\begin{equation}
    \begin{aligned}
        N_1 \equiv N_{c_1, \sigma_1} = N_{\sigma_1}, \\
        N_2 \equiv N_{c_2, \sigma_2} = N_{\sigma_2}.
    \end{aligned}
\end{equation}
This implies that the $M_{i,j}$ can be summarized with just two values, each
corresponding to a class:
\begin{equation}
    \begin{aligned}
        M_1 \equiv M_{1, 2} = 1 - M_{1, 1}, \\
        M_2 \equiv M_{2, 1} = 1 - M_{2, 2}.
    \end{aligned}
\end{equation}
Let us consider two varieties $1$ and $2$. This could be the use of standard
language ($1$ means they do, $2$ means they do not). Now let us introduce an
intrinsic prestige $s$ for the variety $2$, such that:
\begin{equation}
\label{eq:lv_intro}
    \begin{aligned}
        P(1 \rightarrow 2) &\propto s, \\
        P(2 \rightarrow 1) &\propto 1 - s.
    \end{aligned}
\end{equation}
Without loss of generality, let us assume $s > 1/2$, meaning $2$ is more prestigious
than $1$. And let us introduce an asymmetric attachment of each group for their own
variety, $q_1$ and $q_2$:
\begin{equation}
\label{eq:qs_intro}
    \begin{aligned}
        P(2 \rightarrow 1 \mid \sigma = \sigma_1) &\propto q_1 > 1/2, \quad&  P(1 \rightarrow 2 \mid \sigma = \sigma_1) \propto 1 - q_1, \\
        P(1 \rightarrow 2 \mid \sigma = \sigma_2) &\propto q_2, \quad&  P(2 \rightarrow 1 \mid \sigma = \sigma_2) \propto 1 - q_2.
    \end{aligned}
\end{equation}
So $2$ is more prestigious than $1$, but individuals of class $\sigma_1$ prefer $1$.

We will write $p_1$ the proportion of individuals of class 1 speaking non-standard (variety 1), and $p_2$ the proportion of individuals of class 2 speaking standard (variety 2).
Then, working in mean-field, one can write the following master equations:
\begin{equation}
    \label{eq:time_evol}
    \begin{aligned}
        % v and c depend on t
        \dv{p_1}{t}
            &= (1 - p_1) P(2 \rightarrow 1 \mid \sigma = \sigma_1)
                - p_1 P(1 \rightarrow 2 \mid \sigma = \sigma_1)
        \\
        \dv{p_2}{t}
            &= (1 - p_2) P(1 \rightarrow 2 \mid \sigma = \sigma_2)
                 - p_2 P(2 \rightarrow 1 \mid \sigma = \sigma_2)
    \end{aligned}
\end{equation}

\subsection{Deriving the master equations}
Following the definitions in \cref{eq:lv_intro,eq:qs_intro} of the influence of $s$, $q_1$ and $q_2$, the transition probabilities can be written as follows:
\begin{equation}
    \label{eq:first_trans_probs}
    \begin{aligned}
        % v and c depend on t
        P(1 \rightarrow 2 \mid \sigma = \sigma_1, c_t = c_j)
            &= s (1 - q_1) P(v_{t-1} = 2 \mid c_t = c_j)
        \\
        P(1 \rightarrow 2 \mid \sigma = \sigma_2, c_t = c_j)
            &= s q_2 P(v_{t-1} = 2 \mid c_t = c_j)
        \\
        P(2 \rightarrow 1 \mid \sigma = \sigma_1, c_t = c_j)
            &= (1 - s) q_1 P(v_{t-1} = 1 \mid c_t = c_j)
        \\
        P(2 \rightarrow 1 \mid \sigma = \sigma_2, c_t = c_j)
            &= (1 - s) (1 - q_2) P(v_{t-1} = 1 \mid c_t = c_j)
    \end{aligned}
\end{equation}
with $P(1 \rightarrow 2) \equiv P(v_t = 2 \mid v_{t-1} = 1)$, $P(2 \rightarrow 1) \equiv
P(v_t = 1 \mid v_{t-1} = 2)$, $t$ denoting the current time step, $P(v_{t-1} = v \mid
c_t = c_j)$ the probability to pick an individual who used variety $v$ at $t-1$ and who is at
cell $c_j$ at $t$. Decomposing it by SE class, and using Bayes' rule, we get:
\begin{equation}
    \begin{aligned}
        &P(v_{t-1} = v \mid c_t = c_j)
        \\[1ex]
        &
        \begin{aligned}
            & = \sum_{k} P(v_{t-1} = v, \sigma = \sigma_k \mid c_t = c_j) \\
            &= \sum_{k} P(v_{t-1} = v, \sigma = \sigma_k)
                \cdot \frac{P(c_t = c_j \mid v_{t-1} = v, \sigma = \sigma_k)}{P(c_t = c_j)} \\
            &= \sum_{k} P(v_{t-1} = v \mid \sigma = \sigma_k)
                \cdot P(\sigma = \sigma_k)
                \cdot \frac{P(c_t = c_j \mid v_{t-1} = v, \sigma = \sigma_k)}{P(c_t = c_j)}. \\
        \end{aligned}
    \end{aligned}
\end{equation}
Let us introduce $p_{v, \sigma_k} \equiv P(v_{t-1} = v \mid \sigma = \sigma_k)$ to simplify equations
further, which summarise the state of the system at the previous step. Also, since
they satisfy $\sum_v p_{v, \sigma_k} = 1$, we will only write in terms of $p_1 \equiv p_{1,
1}$ and $p_2 \equiv p_{2, 2}$. We also have $P(\sigma = \sigma_k) = \frac{N_{\sigma_k}}{\sum_k
N_{\sigma_k}}$. The final term in the product above is related to the mobility of the different SES
classes. Indeed, aligning the indices of the SE class with the one of their cell of
residence, and using the fact that the random variable $C_{t}$, which represents the
cell where an individual will be encountered at step $t$, is independent from $V_{t-1}$,
which represents an individual's variety usage at the previous step, we have:
\begin{equation}
\label{eq:mob_by_SES}
    P(c_t = c_j \mid v_{t-1} = v, \sigma = \sigma_k) = P(c = c_j \mid \sigma = \sigma_k) = M_{k, j},
\end{equation}
and
\begin{equation}
    P(c_t = c_j) = \sum_k P(c = c_j \mid \sigma = \sigma_k) P(\sigma = \sigma_k) = \sum_k M_{k, j} \frac{N_{\sigma_k}}{\sum_k
    N_{\sigma_k}}.
\end{equation}
% and $\sum_j m_{k, j} = 1$.
Let us now introduce
\begin{equation}
    m_{k, j} \equiv \frac{N_{\sigma_k} M_{k, j}}{\sum_{k'} N_{\sigma_{k'}} M_{k', j}},
\end{equation}
which satisfy $\sum_k m_{k, j} = 1$ (unlike the $M_{k, j}$ that satisfy $\sum_j M_{k, j}
= 1$). It is simply the expected proportion of individuals that will appear in $j$ that
have status $k$. We can then write
\begin{equation}
\label{eq:decomp_prev_by_SES}
    P(v_{t-1} = v \mid c_t = c_j) = \sum_k p_{v, \sigma_k} m_{k, j}.
\end{equation}

We will further abuse notation and only use $m_1 \equiv
m_{1,2}$ and $m_2 \equiv m_{2, 1}$, which summarize how mobile each group is. Let us now rewrite
\eqref{eq:decomp_prev_by_SES} in terms of $m_1$, $m_2$, $p_1$ and $p_2$ only:
\begin{equation}
    \label{eq:prev_step_mob}
    \begin{aligned}
        P(v_{t-1} = 2 \mid c = c_1)
            = (1 - m_2) (1 - p_1) + m_2 p_2
        \\
        P(v_{t-1} = 2 \mid c = c_2)
            = m_1 (1 - p_1) + (1 - m_1) p_2
        \\
        P(v_{t-1} = 1 \mid c = c_1)
            = (1 - m_2) p_1 + m_2 (1 - p_2)
        \\
        P(v_{t-1} = 1 \mid c = c_2)
            = m_1 p_1 + (1 - m_1) (1 - p_2)
    \end{aligned}
\end{equation}
We can subsequently write the forms in \eqref{eq:first_trans_probs} in terms of these four
variables only. But what we actually want is to write $P(v_t = v \mid v_{t-1} = v', \sigma
= \sigma_k)$ for $v' \neq v$. Decomposing this one by cell, we can get the following:
\begin{equation}
    \label{eq:trans_probs_cell_decomp}
    \begin{aligned}
        & P(v_t = v \mid v_{t-1} = v', \sigma = \sigma_k)
        \\[1ex]
        &
        \begin{aligned}
            \quad & = \sum_j P(c = c_j \mid \sigma = \sigma_k)
            \cdot P(v_t = v \mid v_{t-1} = v', \sigma = \sigma_k, c_t = c_j)
        \\
            & = \sum_j M_{k, j} P(v_t = v \mid v_{t-1} = v', \sigma = \sigma_k, c_t = c_j).
        \end{aligned}
    \end{aligned}
\end{equation}
Finally, inserting \eqref{eq:first_trans_probs} into \eqref{eq:trans_probs_cell_decomp}, we get:
\begin{equation}
    \label{eq:final_trans_probs}
    \begin{aligned}
        P(1 \rightarrow 2 \mid \sigma = \sigma_1)
            &= s (1 - q_1)
            \begin{aligned}[t]
            &[(1 - M_1) P(v_{t-1} = 2 \mid c = c_1)
            \\
            & + M_1 P(v_{t-1} = 2 \mid c = c_2)]
            \end{aligned}
        \\
        P(1 \rightarrow 2 \mid \sigma = \sigma_2)
            &= s q_2
            \begin{aligned}[t]
                &[M_2 P(v_{t-1} = 2 \mid c = c_1)
                \\
                &+ (1 - M_2) P(v_{t-1} = 2 \mid c = c_2)]
            \end{aligned}
        \\
        P(2 \rightarrow 1 \mid \sigma = \sigma_1)
            &= (1 - s) q_1
            \begin{aligned}[t]
                &[(1 - M_1) P(v_{t-1} = 1 \mid c = c_1)
                \\
                &+ M_1 P(v_{t-1} = 1 \mid c = c_2)]
            \end{aligned}
        \\
        P(2 \rightarrow 1 \mid \sigma = \sigma_2)
            &= (1 - s) (1 - q_2)
            \begin{aligned}[t]
                &[M_2 P(v_{t-1} = 1 \mid c = c_1)
                \\
                &+ (1 - M_2) P(v_{t-1} = 1 \mid c = c_2)]
            \end{aligned}
    \end{aligned}
\end{equation}

% Note: the mobility matrix' Pearson r in this case is equal to $r_M = 1 - M_1 - M_2$.

\subsection{Case of equal populations and mobility}
In the following, we will assume $N_{\sigma_1} = N_{\sigma_2}$, which implies that $m_{k, j} =
\frac{M_{k, j}}{\sum_{k'} M_{k', j}}$. If we assume equal mobility, introducing $M \equiv
M_1 = M_2$, we have $m_1 = m_2 = M$, and it follows that:
\begin{equation}
    \begin{aligned}
        P(1 \rightarrow 2 \mid \sigma = \sigma_1)
            &= s (1 - q_1)
            \left[
              M^* (p_1 + p_2 - 1) + 1 - p_1
            \right]
        \\
        P(1 \rightarrow 2 \mid \sigma = \sigma_2)
            &= s q_2
            \left[
              M^* (1 - p_1 - p_2) + p_2
            \right]
        \\
        P(2 \rightarrow 1 \mid \sigma = \sigma_1)
            &= (1 - s) q_1
            \left[
              M^* (1 - p_1 - p_2) + p_1
            \right]
        \\
        P(2 \rightarrow 1 \mid \sigma = \sigma_2)
            &= (1 - s) (1 - q_2)
            \left[
              M^* (p_1 + p_2 - 1) + 1 - p_2
            \right]
    \end{aligned}
\end{equation}
with $M^* = 2 M (1 - M)$. We thus get to the result presented in the main text:
\begin{equation}
  \label{eq:ses_ling_time_evol_eq_mob}
  \left\{
  \begin{aligned}
      \dv{p_1}{t}
          &=
            2 M (1 - M) (1 - p_1 - p_2) [q_1 (1 - s) - p_1 (q_1 - s)]
      \\
          & \quad +
            p_1 (1 - p_1) (q_1 - s)
      \\[1ex]
      \dv{p_2}{t}
          &=
            2 M (1 - M) (1 - p_1 - p_2) [q_2 s - p_2 (s + q_2 - 1)]
      \\
          & \quad +
            p_2 (1 - p_2) (s + q_2 - 1)
  \end{aligned}
  \right.
\end{equation}

\subsection{Coexistence solution}
\label{sec:coex_sol_ses_model}
Let us assume there exists a fixed point of \cref{eq:ses_ling_time_evol_eq_mob}, denoted
$(p_1^*, p_2^*)$, which is such that $0 < p_1^* < 1$ and $0 < p_2^* < 1$. This would
correspond to a state of coexistence of the two varieties within both classes. We want here to find out
under what conditions the existence of such a fixed point is not possible. We assume all
parameters of the system are in the open unit interval. We found through symbolic
computations that $q_1 \leq s$ prohibits its existence, but let us prove it by
contradiction in a simple case, which is here the most physically relevant.

First, for $q_1 = s$:
\begin{equation}
    \dv{p_1}{t} = 0 \Rightarrow p_1^* = 1 - p_2^*.
\end{equation}
It directly follows from the condition $\dv{p_2}{t} = 0$ that $p_2^*$ must be
either $0$ or $1$. In this case, there is therefore no possible coexistence.

Let us then assume $q_1 < s$. We have
\begin{equation}
    \label{eq:appendix_eq_points}
    \begin{aligned}
        & \dv{p_1}{t} = 0
        \\[1ex]
        & \Rightarrow 2 M (1 - M) (1 - p_1^* - p_2^*) \left[q_1 \frac{1 - s}{s - q_1} + p_1^* \right] = p_1^* (1 - p_1^*).
    \end{aligned}
\end{equation}
As $s - q_1 > 0$ and $0 < p_1^* < 1$, both the right-hand side and the term in square
bracket of the left-hand side are strictly positive. But we also have
\begin{equation}
    \begin{aligned}
        & \dv{p_2}{t} = 0
        \\[1ex]
        & \Rightarrow 2 M (1 - M) (1 - p_1^* - p_2^*)
            = - \frac{p_2^* (1 - p_2^*) (s + q_2 - 1)}{q_2 s - p_2^* (s + q_2 - 1)}.
    \end{aligned}
\end{equation}
Now, let us assume that $s + q_2 - 1 > 0$. It is the case that makes sense here,
since variety 2, corresponding to standard language, has a higher prestige than 1, so
$s > 0.5$, and at least a neutral bias of the high SE class, so $q_2 \geq 0.5$.
Since $0 < p_2^* < 1$, this implies that the right-hand side above is negative, hence
\begin{equation}
    2 M (1 - M) (1 - p_1^* - p_2^*) < 0.
\end{equation}
This is in contradiction with the signs of the other terms in
\cref{eq:appendix_eq_points}. It is therefore strictly impossible to have coexistence
solutions when $q_1 < s$ and $s + q_2 - 1 > 0$. In other words, for the non-standard form to survive, it is necessary that the
SE class 1 has a positive bias towards the non-standard variety that is higher than the
prestige of the standard form.

\clearpage

\clearpage

\begin{table}[p]
  \centering
  \begin{filecontents*}{data/cats_global_stats.csv}
name,r_with_net_income18,count
Grammar,-0.25004387074552137,1693408
Commonly Confused Words,-0.20781308418966482,283731
Possible Typo,-0.20258387122352264,1977466
Miscellaneous,-0.12025789232489195,790404
Punctuation,-0.08962916357656203,3525135
Style,-0.07190251849105202,708659
Compounding,-0.0448003265219161,88391
Collocations,-0.04045727580825972,46814
Nonstandard Phrases,-0.0027528938334869706,5439
Redundant Phrases,0.018145310250248257,103643
Semantics,0.03248584613852084,6425
Repetitions (Style),0.037925894053217224,1741
Upper/Lowercase,0.10703396832462368,2854732
Typography,0.2125337285012526,9357945
\end{filecontents*}
\pgfplotstabletypeset[
    string type,
    columns={name,count,r_with_net_income18},
    every head row/.style={before row=\toprule, after row=\midrule},
    every last row/.style={after row=\bottomrule},
    columns/name/.style={column type=l, column name={Rule category}},
    columns/count/.style={column type={S[table-format=7.0]}, column name=\text{Count}},
    columns/r_with_net_income18/.style={
      column type={S[table-auto-round, table-format=1.2]},
      column name=\text{Pearson r correlation with net income},
    },
  ]
  {data/cats_global_stats.csv}
  \caption[Summary statistics about each category of rules defined by LanguageTool.]{\textbf{Summary statistics about each category of rules defined by LanguageTool.} The number of detected deviations from rules, as well as the Pearson r correlation with the net income in the \SI{4879}{} MSOAs left after our filters.}
  \label{tab:cats_global_stats}
\end{table}

\begin{table}[p]
  \centering
  \begin{filecontents*}{data/top_grammar_mistakes.csv}
rule_id,name,count
HE_VERB_AGR,Agreement error: Non-third person/past tense verb with 'he/she/it' or a pronoun,122203
THE_SUPERLATIVE,Zero or indefinite article ('a'/'an') before superlatives,69461
PRP_VBG,He going (He is going),58775
CD_NN,Possible agreement error: numeral + singular countable noun,40411
ITS_TO_IT_S,I have to do laundry while its (it's) still sunny,33667
PHRASE_REPETITION,Repetition of two words ('at the at the'),33414
PRP_PAST_PART,Agreement error: past participle without 'have',32913
A_NNS,Agreement: 'a' + plural word,32902
BEEN_PART_AGREEMENT,Agreement: 'been' or 'was' + past tense,31697
CAUSE_BECAUSE,confusion of cause vs. because,30511
\end{filecontents*}
\pgfplotstabletypeset[
    string type,
    columns={rule_id,name,count},
    every head row/.style={before row=\toprule, after row=\midrule},
    every last row/.style={after row=\bottomrule},
    columns/rule_id/.style={
      verb string type,
      column type=l,
      column name={Rule ID},
      postproc cell content/.code={%
        \pgfplotsutilstrreplace{_}{\_}{##1}%
        \pgfkeyslet{/pgfplots/table/@cell content}\pgfplotsretval
      },
    },
    columns/name/.style={column type={p{8cm}}, column name={Description}},
    columns/count/.style={column type={S[table-format=7.0]}, column name=\text{Count}},
  ]
  {data/top_grammar_mistakes.csv}
  \caption[Ten most frequently detected grammar rules.]{\textbf{Ten most frequently detected grammar rules.} For each of them, we provide the rule ID assigned by LanguageTool, its description and their number of occurrences identified in the tweets of our home-located Twitter users.}
  \label{tab:top_grammar_mistakes}
\end{table}

\begin{table}[p]
  \centering
  \begin{filecontents*}{data/uk_metropolitan_areas.csv}
FID,LAD20CD,LAD20NM,CAUTH20CD,metro_area
1,E08000001,Bolton,E47000001,Manchester
2,E08000002,Bury,E47000001,Manchester
3,E08000003,Manchester,E47000001,Manchester
4,E08000004,Oldham,E47000001,Manchester
5,E08000005,Rochdale,E47000001,Manchester
6,E08000006,Salford,E47000001,Manchester
7,E08000007,Stockport,E47000001,Manchester
8,E08000008,Tameside,E47000001,Manchester
9,E08000009,Trafford,E47000001,Manchester
13,E08000018,Rotherham,E47000002,Sheffield
14,E08000019,Sheffield,E47000002,Sheffield
15,E08000032,Bradford,E47000003,Leeds
18,E08000035,Leeds,E47000003,Leeds
20,E06000006,Halton,E47000004,Liverpool
21,E08000011,Knowsley,E47000004,Liverpool
22,E08000012,Liverpool,E47000004,Liverpool
25,E08000015,Wirral,E47000004,Liverpool
31,E08000025,Birmingham,E47000007,Birmingham
33,E08000027,Dudley,E47000007,Birmingham
34,E08000028,Sandwell,E47000007,Birmingham
36,E08000030,Walsall,E47000007,Birmingham
37,E08000031,Wolverhampton,E47000007,Birmingham
45,E06000023,"Bristol, City of",E47000009,"Bristol, City of"
48,E08000023,South Tyneside,E47000010,Newcastle upon Tyne
49,E08000024,Sunderland,E47000010,Newcastle upon Tyne
50,E08000037,Gateshead,E47000010,Newcastle upon Tyne
52,E08000021,Newcastle upon Tyne,E47000011,Newcastle upon Tyne
53,E08000022,North Tyneside,E47000011,Newcastle upon Tyne
\end{filecontents*}
\pgfplotstabletypeset[
    string type,
    text indicator=",
    columns={metro_area,LAD20CD,LAD20NM},
    every head row/.style={before row=\toprule, after row=\midrule},
    every last row/.style={after row=\bottomrule},
    columns/metro_area/.style={column type=l, column name={Metropolitan area}},
  ]
  {data/uk_metropolitan_areas.csv}
  \caption[Definition of the metropolitan areas used in this study]{\textbf{Definition of the metropolitan areas used in this study}. For each metropolitan area, the code and the name of the local authority districts (LADs) contained within each area are given. London is defined as the London region (identified by the code E12000007).}
  \label{tab:uk_metropolitan_areas}
\end{table}

\begin{table}[p]
  \centering
  \begin{filecontents*}{data/dataset_summary.csv}
subreg,avg_nr_tweets,avg_nr_words,avg_nr_unique_words,sum_nr_users,sum_nr_tweets,sum_nr_words,nr_users,uavg_grammar_freq
Birmingham,275.9323792486583,3801.3706618962433,726.6289803220036,2599,770000,10613724,2599,0.0037070342620457437
"Bristol, City of",243.89889106327462,3407.9458577951727,705.8988910632746,1437,373245,5218801,1437,0.0033058572447337577
Leeds,262.22105263157897,3710.9139001349527,716.8304993252361,3393,969553,13731825,3393,0.0035237158791041156
Liverpool,287.08929836995037,3954.619891330026,715.7287975431136,3910,1213219,16721889,3910,0.003806536247086603
London,264.07250134096194,3741.029992848203,702.0754961559091,20469,5896489,83594913,20469,0.003239904403125136
Manchester,288.4792339640055,4084.714928472543,746.1188278726349,8042,2496379,35370104,8042,0.0036253209604693434
Newcastle upon Tyne,287.8534718425369,4019.9682886823402,741.0057408419901,3404,1051460,14691320,3404,0.0038919626502280933
Sheffield,287.9248259860789,4145.303480278422,742.2872389791183,2013,619512,8924594,2013,0.0036591729236621004
other,260.3356013738258,3599.5563777804264,690.9539746687325,86135,23986776,331863788,86135,0.0035180749042406766
England and Wales,264.76434774000114,3686.3234287492223,700.4582649770889,131402,37376633,520730958,131402,0.003503299765884766
\end{filecontents*}
\pgfplotstabletypeset[
    text indicator=",
    columns={subreg, sum_nr_tweets, avg_nr_tweets, sum_nr_words, avg_nr_words, uavg_grammar_freq},
    skip rows between index={8}{9},
    string type,
    column type=S,
    every head row/.style={
      output empty row,
      before row={
        \toprule
        & \multicolumn{2}{c}{Tweets} & \multicolumn{2}{c}{Tokens} & {Deviations per token}\\
        \cmidrule(lr{.5em}){2-3} \cmidrule(lr{.5em}){4-5}
        \multirow{-2}{*}{Location} & {Sum} & {Per user} & {Sum} & {Per user} & {User average} \\
      },
      after row=\midrule
    },
    every last row/.style={after row=\bottomrule},
    columns/subreg/.style={column type=l},
    columns/sum_nr_tweets/.style={column type={S[table-format=8.0]}},
    columns/avg_nr_tweets/.style={
      column type={S[table-auto-round, table-format=3.1]}
    },
    columns/sum_nr_words/.style={column type={S[table-format=9.0]}},
    columns/avg_nr_words/.style={
      column type={S[table-auto-round, table-format=4.1]}
    },
    columns/uavg_grammar_freq/.style={column type={S[exponent-mode=fixed, fixed-exponent=-3, table-auto-round, table-format=1.2e1]}}
  ]
  {data/dataset_summary.csv}
  \caption[Summary statistics of our Twitter corpus.]{\textbf{Summary statistics of our Twitter corpus.} The number of tweets and tokens and their average number per user, as well as the frequency of non-standard features averaged over users are given for our eight metropolitan areas and all of England and Wales.}
  \label{tab:dataset_summary}
\end{table}

\begin{table}
  \begin{tabular*}{.9\textwidth}{l c c c}
    \toprule
    & \multicolumn{2}{l}{Proportion of multilinguals:} \\ \cmidrule(lr{.5em}){2-4}
    & Average & Correlation with grammar deviations & p-value \\ \midrule
    Birmingham & $3.8 \%$ & $-0.12$ & $0.26$ \\
    Bristol, City of & $4.3 \%$ & $-0.26$ & $0.10$ \\
    Leeds  & $2.9 \%$ & $0.08$ & $0.44$ \\
    Liverpool  & $3.9 \%$ & $0.02$ & $0.83$ \\
    London  & $7.6 \%$ & $0.00$ & $0.94$ \\
    Manchester  & $3.5 \%$ & $0.03$ & $0.67$ \\
    Newcastle upon Tyne  & $4.5 \%$ & $-0.14$ & $0.14$ \\
    Sheffield & $4.2 \%$ & $-0.06$ & $0.62$ \\ \bottomrule
  \end{tabular*}
  \caption[Proportions of multilinguals among our identified residents and their correlation with grammar deviations.]{\textbf{Proportions of multilinguals among our identified residents and their correlation with grammar deviations.}}
  \label{tab:multiling}
\end{table}